\documentclass{aa}
\usepackage{appendix}
\usepackage{natbib}
\usepackage{graphicx}
\usepackage{txfonts}
\usepackage{booktabs}
\usepackage{bm}
\usepackage[normalem]{ulem}
\usepackage{xcolor}
\usepackage{nicefrac}
\usepackage{xspace}
\usepackage{xcolor}
\usepackage{hyperref} 
\hypersetup{
colorlinks=true, 
allcolors= black, 
citecolor=blue
}
\usepackage{tikz}
\definecolor{lime}{HTML}{A6CE39}

\DeclareRobustCommand{\ion}[2]{%
\relax\ifmmode
\ifx\testbx\f@series
{\mathbf{#1\,\mathsc{#2}}}\else
{\mathrm{#1\,\mathsc{#2}}}\fi
\else\textup{#1\,{\mdseries\textsc{#2}}}%
\fi}
\begin{document} 
\def\referee#1{\textcolor{blue}{\textbf{#1}}}                                                  
\def\authors#1{\textcolor{red}{\textbf{#1}}} 
\def\refs#1{\textcolor{green}{\textbf{#1}}} 
\title{Something new under the Sun: A magnetically driven CH/CN anti-correlation}
\titlerunning{Something new under the Sun}
\authorrunning{Y. Momany et al.}
\author{
%
{Y. Momany\inst{1}\thanks{
    On  leave  from  INAF-OAPD  and  presently serves  as Scientific
    Attach\'{e}  at the  Italian  Embassy in  Abu Dhabi (United Arab Emirates).}},
{L. Monaco\inst{2,3}},
{S. Villanova\inst{2}},
{S. Zaggia\inst{1}},
{I. Saviane\inst{4}},
{L. Girardi\inst{1}},
{G. Volpato\inst{5,1}},
{G. Costa\inst{6,1}},
{N.R. Landin\inst{7}},
{I. Yegorova\inst{4}},
{M. Montalto\inst{8}},
{M. Dima\inst{1}},
{F. R. Herpich\inst{9}},
\and
{F. Almeida-Fernandes\inst{10}},
}
\institute{INAF--Osservatorio    Astronomico     di    Padova,    Vic.
  dell'Osservatorio        5,        35122       Padova,        Italy,
  \email{yazan.almomany@inaf.it}
  %
\and
Universidad Andres Bello, Facultad de Ciencias Exactas, Departamento de F\'isica y Astronom\'ia - Instituto de Astrofísica, Autopista
  Concepc\'ion-Talcahuano 7100, Talcahuano, Chile
\and 
INAF--Osservatorio    Astronomico     di  Trieste, Via G.B.Tiepolo 11, 34143, Trieste,  Italy
\and 
European Southern  Observatory,   Alonso  de   Cordova  3107,   Santiago,  Chile
\and 
Institut d'Astronomie et d'Astrophysique, Universit\'e Libre de Bruxelles (ULB), CP 226, 1050 Brussels, Belgium 
\and 
Dipartimento di Fisica e Astronomia Galileo Galilei, Universita di Padova, Vicolo dell'Osservatorio 3, I-35122 Padova, Italy
\and 
Universidade Federal de Vi{\c{c}}osa, Campus UFV Florestal, CEP 35690-000 – Florestal, MG, Brazil
\and 
INAF--Osservatorio Astrofisico di Catania, Via Santa Sofia 78, I-95123 Catania, Italy
\and 
Laborat\'orio Nacional de Astrof\'isica (LNA/MCTI), Rua Estados Unidos, 154, Itajub\'a 37504-364, Brazil
\and 
Instituto Nacional de Pesquisas Espaciais,  Av. dos Astronautas 1758, Jardim da Granja,12227-010 S\~ao Jos\'e dos Campos, SP, Brazil
}
\date{Received 30 May  2026; accepted 9 July  2026}
%
%
\abstract
{
  Spectroscopic monitoring surveys have  shown that the magnetic cycle
  alters  the  photospheric  structure  of  the  Sun,  perturbing  the
  formation of  molecular lines.  In parallel,  the study  of globular
  clusters (GCs)  has remained centred on the  multiple stellar populations
  (MPs) conundrum, 
  a  phenomenon   defined---most  notably---by   the  anti-correlation
  between CH  and CN molecular  bands, serving to  distinguish between
  first-population (1P) and second-population (2P) stars.
}
{  Since these  molecular transitions  are intrinsically  sensitive to
  photospheric  conditions, we  investigated  whether stellar  magnetic
  activity  can induce  spectroscopic variations  consistent with  the
  signatures  traditionally   utilised  to   disentangle  1P   and  2P 
  populations. }
{  We bridged  the  stellar  and solar  domains  by applying  standard
  spectroscopic  index  analyses—routinely  used to  identify  MPs  in
  clusters—to    a    decade-long    dataset    of    disc-integrated 
  high-resolution  solar  spectra.   We systematically  monitored  the
  behaviour of  the CH  and CN  molecular bands as  a function  of the
  solar  magnetic  cycle.  These  observational  results  are  further
  complemented by first-order magnetohydrodynamic (MHD) simulations to
  evaluate  how   these  signatures  scale  with   increased  magnetic
  activity.  }
%
{ We  demonstrate that the  disc-integrated solar spectrum  exhibits a
  distinct CH/CN  anti-correlation driven  by the magnetic  cycle.
  Our  simulations indicate  that these  variations in  molecular line
  formation  scale  significantly   with  increased  magnetic  filling
  factors,  rendering   them,  in   principle,  capable   of  reaching
  amplitudes comparable to the  spectroscopic spreads observed between
  1P and 2P stars in GCs. }
%
{  While  the  broader  implications  of  our  discovery  of  a  CH/CN
  anti-correlation  in   the  single-star  Sun  remain   to  be  fully
  elucidated,  our   results  suggest   that  the   spectroscopic  and
  photometric anomalies  defining the  MP conundrum may  be physically
  linked to  a differential  spectral response  to varying  degrees of
  surface  magnetic coverage.
  The emerging  mechanism depends on  the intrinsic magnetic  state of
  the   star---distinguishing   relatively  active   from   relatively
  quiescent  stars---rather than  the  existence of  a magnetic  cycle
  itself.
  Surface magnetism offers a novel physical framework that may provide
  new  insights into  the origin  of MPs  and clearly  warrants future
  exploration.
}
\maketitle%
\section{Introduction}
\label{s_intro}

It is now a well-established  realisation that the ultimate barrier to
detecting and characterising low-mass  Earth-analogue exoplanets is no
longer set by our instrumental precision,  but rather by the host star
being investigated (e.g. \citealt{fischer2016,pepe2014}).
Indeed,  intrinsic   stellar  variability,  driven  by   the  restless
interplay  of magnetic  activity  and  surface convection, introduces
spurious Doppler shifts and spectral line deformations that can easily
mask or  mimic delicate  planetary signals.   Faced with  the daunting
task of  decoding these complex  spectral signatures, it is  no wonder
that astronomers naturally  returned their attention to  the only star
whose surface  can be  spatially resolved  and monitored  in exquisite
detail: our Sun.

Historically,     a     profound    disconnect     existed     between
high-spatial-resolution   solar  physics   and  the   disc-integrated 
point-source observations typical of  stellar astrophysics.
Although  solar physicists  could  meticulously map  the evolution  of
individual   sunspots,  faculae,   and   granulation  cells,   stellar
astronomers  could  only  observe  the net  diluted  effect  of  these
features mixed into a single unresolved spectrum.
Bridging  this fundamental  gap required  a shift  in perspective:  We
needed to observe the Sun exactly as we observed distant stars.

This  necessity gave  rise  to  the now-flourishing  `Sun-as-a-star'  
paradigm. Although this approach  has  recently  experienced a  vibrant
renaissance, its foundational pillars were erected decades ago.
We owe a great deal to  the pioneering foresight of William Livingston
and collaborators, who at the McMath-Pierce Solar Telescope (Kitt Peak
National Observatory), dedicated over  three decades to systematically
monitoring  the  integrated solar  disc.
Their enduring  baseline of observations, initiated  in the mid-1970s,
established   the   absolute    bedrock   of   Sun-as-a-star   studies
\citep[e.g.][]{white1981,   livingston2007},   providing  the   vital
empirical  link  between resolved  solar  surface  features and  their
integrated spectroscopic signatures.
Several decades before the  modern exoplanet era took  hold, they definitively
demonstrated  how the  solar  cycle subtly  but persistently  imprints
itself on spectral line parameters.

Building on  this rich historical  legacy, the  modern era has  seen a
proliferation of global initiatives that  aim to push this paradigm to
unprecedented levels of  precision, creating a network  of solar feeds
connected          to           next-generation          spectrographs
(\citealt{phillips2016,lin2022}). Among  these initiatives,  a notable
milestone was achieved by coupling  a small dedicated solar telescope
to  the  HARPS-N  spectrograph  at the  Telescopio  Nazionale  Galileo
\citep{dumusque2015}.   This   innovative   setup  allowed   for   the
continuous high-cadence acquisition  of disc-integrated solar spectra
using  the  very  same ultra-stable,  high-resolution  instrumentation
deployed for cutting-edge exoplanet searches.

The  insights drawn  from  these modern  Sun-as-a-star campaigns  have
painted a remarkably detailed picture  of stellar variability across a
multitude  of timescales.
On the  shortest timescales of  minutes to hours,  the disc-integrated
spectrum is  dominated by the  acoustic ringing of  solar oscillations
and the  boiling surface of granulation  \citep{meunier2010}.
Over days to weeks, the rotation  of the Sun steps into the spotlight,
modulating the visibility of active regions. Most interestingly, these
early  high-cadence  studies  definitively   showed  that  for  slowly
rotating, relatively inactive stars such as  the Sun, the dominant driver
of  radial  velocity  variation   is  the  suppression  of  convective
blue shift  by  bright,  magnetic   faculae,  rather  than  the  simple
flux-blocking         effect          of         dark         sunspots
(\citealt{haywood2016,milbourne2019}).

Today, these  continuous, high-fidelity surveys have  matured to reach
critical,  multi-decade milestones  (\citealt{dumusque2025}), granting
us  a pristine  window into  global stellar  dynamos and  the magnetic
activity cycle itself.
Consequently,  the Sun-as-a-star  framework is  undergoing a  profound
conceptual shift.   Originally conceived  primarily as  an engineering
tool to calibrate out exoplanet  noise, it can organically evolve into
an unparalleled laboratory for fundamental stellar astrophysics.

It is now abundantly clear that  over an $11$-year cycle, the evolving
magnetic network  subtly alters the global  thermodynamic structure of
the stellar  photosphere, inducing long-term shifts  in the integrated
spectrum (\citealt{meunier2017}).
Quite simply, because the physical  conditions of the photosphere vary
with the magnetic cycle, the  formation depths and opacities of atomic
and  molecular lines  are  inevitably perturbed  (\citealt{wise2022}).
Recognising  this,  researchers  are   looking  beyond
standard radial  velocities to investigate how  magnetic cycles deeply
influence  individual spectral  line  depths,  widths, bisectors,  and
importantly,         chemical         abundance         determinations
(\citealt{pietrow2026}).

In a remarkably different domain of stellar astrophysics, the study of
globular clusters (GCs) has undergone  a notable change.  For decades,
GCs  were considered  the quintessential  examples of  `simple stellar
populations'---vast assemblages  of stars born  at the same  time from
the  same   molecular  cloud,   sharing  identical   initial  chemical
compositions.   This  paradigm  provided the  foundation  for  stellar
evolution  theory and  the  dating  of the  oldest  structures in  the
Universe \citep[e.g.][]{gratton2004}.

However,  high-resolution spectroscopy  and high-precision  photometry
have altered this classical view. Extensive surveys have revealed that
virtually  all  old,  massive  GCs  host  multiple  populations  (MPs)
characterised by distinctive, star-to-star variations in light-element
abundances \citep{gratton2012, piotto2015}.  The landmark Hubble Space
Telescope  UV  Legacy  Survey unequivocally  demonstrated  that  these
chemical variations translate into  complex, discrete sequences across
the  colour-magnitude  diagrams  of  GCs,   proving  that  MPs  are  a
ubiquitous  phenomenon in  these  ancient systems  \citep{renzini2015,
  milone2017}.

Interestingly,  decades before  these  discrete photometric  sequences
were mapped, the underlying spectroscopic footprint of this phenomenon
had already been recognised as early  as 1979 through the discovery of
striking          chemical           anti-correlations          (e.g. 
\citealt{kraft1979,smith1987,briley2004a}).
Within a  single cluster, stars exhibit  anti-correlated abundances of
carbon and nitrogen, sodium and oxygen, and occasionally magnesium and
aluminium  \citep{bastian2018}.   Because carbon and  nitrogen variations
are particularly pronounced,  molecular bands such as  CH (the G band)
and CN have  become the primary, most  sensitive observational tracers
for distinguishing between first-population (1P) and second-population
(2P) cluster stars.

Despite  overwhelming  observational  evidence, the  origin  of  these
chemical anomalies remains one of  the greatest unsolved conundrums in
modern  astrophysics. The  presence of  hot-hydrogen burning  products
(e.g. enhanced N and Na, depleted C and O) in unevolved main-sequence
stars implies that  the cluster forming environment was  polluted by a
prior generation  of more massive  stars.  Yet all  proposed polluter
candidates   (ranging   from   asymptotic  giant   branch   stars   to
fast-rotating   massive  stars   and  supermassive   stars)  fail   to
simultaneously satisfy  the complex nucleosynthetic,  mass-budget, and
dynamical    constraints    observed     in    present-day    clusters
\citep{cassisi2020}.

These   two  vibrant   domains   of   research  (ultra-precise   solar
spectroscopy  and the  chemical analysis  of ancient  GCs) rarely,  if
ever, overlap in the literature.
However, the  very molecular features utilised  to distinguish between
1P  and 2P  stars,  most notably  the  CH and  CN  bands, are  readily
accessible  in  continuous   disc-integrated  solar  monitoring  data.
Crucially, these  molecular transitions are known  to be intrinsically
sensitive  to  the  evolving   thermodynamic  structure  of  the  solar
photosphere.
Given  the  profound  theoretical  impasse  surrounding  the  globular
cluster  puzzle,  we  decided  to  apply  standard  chemical  analysis
techniques of stellar clusters  to the exquisite, decade-long datasets
of the Sun-as-a-star.
Remarkably, we observed a  distinct empirical parallel.  Mimicking the
spectroscopic signatures often used to disentangle 1P from 2P stars in
GCs,  the  disc-integrated  solar  spectrum   exhibits  a  CH  and  CN
anti-correlation that is driven by the solar magnetic cycle.
While  the true  implications of  this  discovery for  the broader  MP
conundrum remain  to be  fully understood, this  paper aims  simply to
introduce this empirical cycle-driven molecular behaviour. Ultimately,
our hope is to attract the  attention of these two disparate fields of
astrophysics  to  suggest  the  possibility of  a  connection  between
stellar magnetism and  the GC MP puzzle as an  avenue worthy of future
exploration.

The paper is organised as follows.
In Sect.\,\ref{s_data}, we describe the Sun-as-a-star datasets and the
methodology used for their compilation.
Section\,\ref{s_result_solar_chcn}  presents   the  primary  empirical
finding of  this work: the  detection of an  unambiguous cycle-driven
CH/CN anti-correlation in the Sun.
This        result is        then        extended        in
Sect.\,\ref{s_faculae_vs_globulars}, where  we explore the  scaling of
magnetic surface features and establish a viable observational link to
the  anomalous  spectroscopic  properties  of  2P  stars  in  globular
clusters.
Finally, Sect.\,\ref{s_conclusion}  offers our concluding  remarks and
discusses the broader implications of  stellar magnetism for the study
of MPs.

\section{Photometric and spectroscopic datasets}
\label{s_data}

\subsection{Solar irradiance reconstructions from SATIRE-S}

The  primary objective  of  this paper  was to  extract  and analyse  a
continuous  time series  of  solar CH  (methylidyne,  CH4300) and  CN
(cyanogen, violet  CN4142) indices spanning an  entire magnetic cycle.
To achieve this, we utilised the SATIRE-S model \citep{yeo2014}.
The model  provides a  reconstruction of  the daily  solar irradiance,
which  in turn  draws on  full-disc observations  from various  (KPVT,
SoHO/MDI, and SDO/HMI)  space missions, spanning the  period from 1974
to 2013.
Panels~b and  c of Fig.~\ref{f_1_chcn_anticor}  display the
corresponding measurement of CN and CH solar indices during the entire
$23^{\mathrm{rd}}$ solar cycle, which began in August $1996$ and ended
in December $2008$ (lasting approximately ${\sim}12.3$ years).
We made sure that the CH and  CN solar indices were computed using the
exact same wavelength intervals employed  in the reduction of both TNG
solar observations  (cf. Sec.\,\ref{s_tng_data})  and the  FLAMES data
for NGC104 stars (cf. Sec.\,\ref{s_flames_data}).
In        particular,         we        used         the        G band
$4300$$-$$4315$\,\AA\, range  to infer the  CH index (often used  as a
proxy of carbon abundance) and the 
$4126$$-$$4214$\,\AA\, range  to infer the  CN index (often used  as a
proxy of nitrogen abundance).
%

%

To  illustrate  to  the  reader the  significantly  different  spatial
coverage  of   the  major   solar  magnetic  surface   components,  in
Fig.~\ref{f_2_fac_spo} we display three snapshots of the Sun's global
surface morphology at the  extremes of its $24^{\mathrm{th}}$ magnetic
cycle (which spanned from December $2008$ to December $2019$).
This  reconstruction was,  once  again, generated  using the  SATIRE-S
model (\citealt{yeo2014}).
SATIRE-S employs full-disc magnetograms  and continuum
images to quantify both the  fractional disc area coverage and spatial
distribution of  various surface components. These  include: the quiet
Sun, sunspot umbrae and penumbrae, faculae, and magnetic network.
Specifically,  sunspot  positions  are extracted  from  HMI  continuum
images,   while  the   facular  distribution   is  derived   from  HMI
magnetograms.

\subsection{HARPS-N high-resolution spectroscopy}
\label{s_tng_data}

Given  the importance of  inferring the  solar CH/CN
anti-correlation from space-based data, we looked to further
validate this discovery using ground-based solar observations.
Obviously,  the  ability  of space-based  telescopes  to   continuously 
monitor the  Sun across  entire magnetic  cycles remains  unmatched by
ground-based facilities.
Nonetheless,  we  made  a  concerted effort  to  identify  a  suitable
ground-based  full-disc   solar  dataset  with sufficiently  high
signal-to-noise ratio (S/N) and long-term, continuous coverage.

Fortunately,  the project  initiated  using  the HARPS-N  spectrograph
(\citealt{dumusque2015,dumusque2021}),   mounted  at   the  Telescopio
Nazionale Galileo  (TNG) in  Spain, aligned  remarkably well  with our
requirements.
The primary goal of this initiative  is to deepen our understanding of
how   magnetic  activity   in  solar-like   stars  affects--and   often
hinders--the detection of Earth-like exoplanets.
Specifically,  the  programme  aims  to directly  correlate  short-  and
long-term variations  in surface  inhomogeneities with changes  in the
Sun's full-disc radial velocity.
Notably,  analysis of  the accumulated  data from  this compact  solar
telescope has  already led  to the  conclusion (\citealt{haywood2016})
that the dominant contributor to the Sun's radial velocity variability
is  magnetic activity  in  the  form of  the  hot  and bright  facular
component (not the dark and cool sunspots component).

The  solar  telescope  used  to  feed  the  HARPS-N  spectrograph  was
specifically designed to  capture light from the full solar  disc in a
uniform and stable manner (\citealt{phillips2016}).
It features a $200$\,mm focal length  and a $3$-inch aperture lens that
focuses  sunlight into  a  $2$-inch diameter  integrating sphere.  The
output  is then  fibre-coupled via  a $300$\,micron  multi-mode optical
fibre to the calibration unit of the HARPS-N spectrograph.
HARPS-N   is  a   fibre-fed,   cross-dispersed  echelle   spectrograph
engineered for high-precision  radial velocity measurements, primarily
aimed  at  exoplanet  detection.  It  operates  at  the  $3.6$-m  TNG
telescope, located at  the Observatorio del Roque de  los Muchachos in
Spain.
The      instrument     delivers      a     spectral      range     of
$390$$-$$690$\,nm   with    an   impressive   resolving    power   of
$R$$=$$115{,}000$.
HARPS-N is simultaneously fed by two fibres (one for the science target
and one  for the reference) allowing  for real-time calibration  of the
solar or stellar light.

The earliest publicly available HARPS-N/TNG solar spectra date back to
the  $29$   July  $2015$,  which,  unfortunately,  is
${\sim}2$\,years   after   the   expected   maximum   of   the   Sun's
$24^{\mathrm{th}}$ magnetic cycle (around $2013$).
Despite this limitation in capturing solar spectra at the cycle's 
`extremes', we show that the  available data are still sufficient
to support our conclusions based on SATIRE-S data.
To construct  statistically robust  and internally  consistent Epoch-I
and Epoch-II datasets (corresponding to  periods of maximum and
 minimum solar  activity) we first selected  all spectra obtained
within a one-month window centred on each epoch. This approach helped
neutralise the effects of varying spot distributions over a full
solar rotation.
In a  second step, we further  refined the selection by  choosing, for
each day, at  least two spectra taken  when the Sun was  at its lowest
airmass. We also  ensured that the final Epoch-I  and Epoch-II samples
had comparable signal-to-noise ratios.
This      process     yielded      an      Epoch-I     dataset      of
$38$          spectra          collected          between         
$29$  July and  $31$ August $2015$.
Similarly,        the         Epoch-II        dataset        comprised
$36$          spectra           obtained          between          
$14$ June and  $15$ July $2018$.

\subsection{The S-PLUS photometric dataset}
\label{s_splus_data}

To prove a workable connection  between solar facular properties and
the 2P populations in globular clusters, we needed wide-field G band
photometry of  a template cluster.
Such  data allow  us to  investigate whether  the primary  photometric
signatures of faculae---specifically, CH depletion and brighter G band
luminosities---are empirically reflected in 2P stars. Panels~a, b, and
c  of Fig.~\ref{f_3_conundrum}  display  colour-magnitude diagrams  of
NGC104 that provide compelling evidence of this potential link.
These  diagrams are  derived from  data acquired  using the  T80-South
telescope   \citep[T80S;][]{mendes2019},   located    at   the   Cerro   Tololo
Inter-American Observatory in Chile.

Our programme (observed under the identifiers CN2018A-76 and CN2018B-11,
P.I.  Monaco, L.)   originally started in March 2018  as a long-period
variability monitoring  programme of $3$ GCs  (NGC104/47Tuc, NGC6752 and
NGC5139/$\omega$Cen).  The  scientific goal  was to take  advantage of
$3$  fundamental  T80S  properties:   (i)  its  impressive  wide-field
(${\sim}1.4^{\circ}   \times   1.4^{\circ}$)    coverage;   (ii)   the
availability of  both the  classical $ugriz$ Sloan  broad-band filters
and a set of $7$ narrow-band Javalambre filters (\citealt{marin2012});
and (iii) its   single $9.2\times\,9.2$k high-efficiency CCD with
a pixel scale of $0\farcs55$.
The pre-reduction  of T80S images  follows the well-tested  recipes of
the     T80S     pipeline    jype\,v0.9.9     \citep[][see
\citealt{mendes2019}   and  \citealt{almeida2022}   for  an   in-depth
discussion of the pipeline  applied to T80S data]{cristobal2014}.  The
astrometry calibration  of each  image is  calculated using  the SCAMP
software (\citealt{bertin2006})  upon J2000  positions from  the 2MASS
input catalogue.
The    astrometric    pipeline    has   been    successfully    tested
(\citealt{almeida2022,mendes2019}) for the release  of the S-PLUS DR1,
DR2, DR3, and DR4.
Further, the  photometric reduction  of the  T80S datasets
 utilised   the    DAOPHOT\,II/ALLFRAME (\citealt{stetson1994})   point spread
function (PSF) fitting code.
The    code     is    inserted     in    our     well-tested    pseudo
photometric-pipeline (\citealt{momany2004}) which combines  the reduction of
both  the   deep and  the  short exposures  and delivers  a
single,  wide-field,   PSF-based  instrumental-magnitudes  photometric
catalogue.
The    absolute   photometric    calibration   utilised    the   S-PLUS
pipeline \citep{almeida2022}.
The  first step  of  this  calibration process  is  to crossmatch  our
instrumental magnitude catalogues with an external reference catalogue.
In this case, the external reference catalogue consists of a combination
of    the     ATLAS    RefCat2 (\citealt{tonry2018})    and     the    GALEX
DR6+7 (\citealt{bianchi2017})  catalogues.  
The  second  step  of  the  calibration pipeline  is  to  perform  the
SED-fitting  process.   It included  a  correction  for the  estimated
extinction (\citealt{schlegel1998}) of  each star, and adopted  the standard
extinction law (\citealt{cardelli1989}) with $R_{V}$$=$$3.1$.
The  synthetic  magnitudes  are  derived  from  a
spectral library (\citealt{coelho2014})  and a $\chi^2$-minimisation
process is  used to infer the  best fitting models to  the $g,r,i,$ and
$z$ magnitudes from the RefCat2 and the near-UV magnitudes from GALEX.
S-PLUS calibrated magnitudes  are inferred from the best  fit for each
source entry, and the mode  of the distribution of differences between
instrumental  and  predicted  magnitudes   is  used  to  estimate  the
calibration zero-point for each filter.  
We take advantage of  the narrow-band filters'  capacity  to better
constrain the models and refine the final zero-points, and perform a 
second  SED-fitting using  the  pre-calibrated  S-PLUS
magnitudes.
For every given filter, the single  derived zero-point is
then applied to the instrumental magnitudes catalogue.
Lastly, following standard procedures, we cleaned the T80-South NGC104
catalogue of  field stars  using a membership  parameter based  on the
proper motion distance  of the cluster's stars from  the average value
obtained from Gaia DR3 (\citealt{Gaia_DR3}).

\subsection{The FLAMES spectroscopic dataset}
\label{s_flames_data}

In panel~d of  Fig.~\ref{f_3_conundrum}, we display the carbon
and nitrogen  abundances of  red horizontal  branch (red-HB)  stars in
NGC104.
When combined  with panel~c of the  same figure,  these data
prove a viable connection  between the defining characteristics of
solar faculae---specifically, CH depletion paired  with brighter
  G band  luminosities---and  the   observational  signatures  of  2P
populations in globular clusters.

The plotted C and N abundances of  red-HB stars in NGC104 are based on
CH  and  CN  index  measurements,  and a  detailed  account  of  the
derivation of these indices is provided in \cite{villanova2025x}.
Besides  red-HB stars,  our \cite{villanova2025x}  spectroscopic study
collected FLAMES/GIRAFFE  at the VLT  (\citealt{pasquini2003}) spectra
of red giant branch (RGB) and asymptotic giant branch (AGB) stars.
The  GIRAFFE data  employed the  LR02  grism setup,  which covers  the
wavelength                        range                        between
$3964$$-$$4567$\,\AA\, and provided a resolution of $R$$=$$6000$.
Overall, the    resulting     averaged-spectra    had    a     dispersion    of
$0.2$\,\AA\, per pixel and a typical S/N between ${\sim}300$$-$$350$.
Thanks  to  a GAIA-based  proper-motion  pre-selection  of our  target
stars,  all red-HB  stars  plotted in  Fig.~\ref{f_3_conundrum} had  an
average                radial               velocity                of
${\sim}$$-16.1$\,km/s in  excellent agreement with mean  radial velocity
of NGC104  ($-18$\,km/s, as provided in  \citealt{harris2010}) and are
therefore firm cluster members.

\section{The discovery: Unveiling a solar CH/CN anti-correlation}
\label{s_result_solar_chcn}

\begin{figure}
\centering
\includegraphics[width=\hsize]{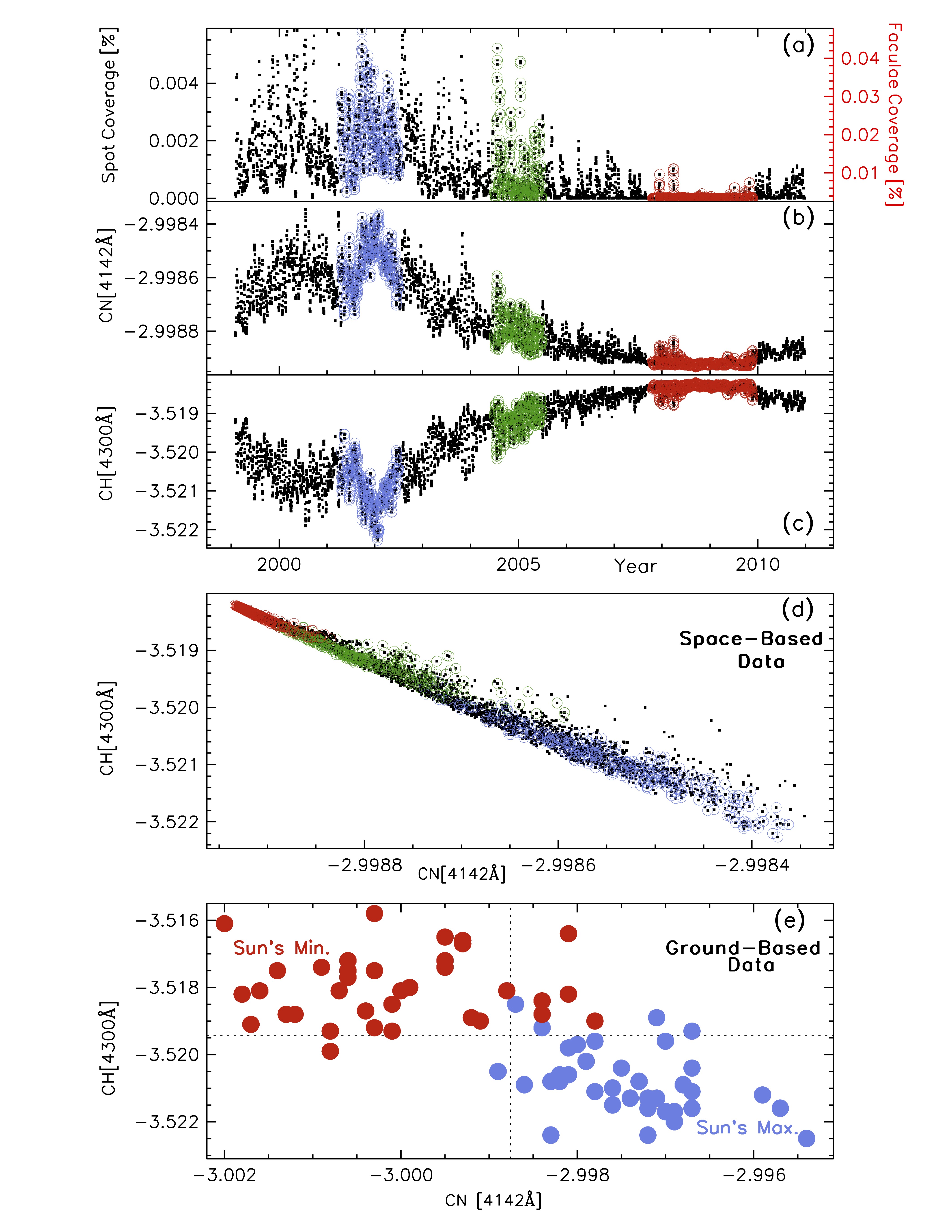} 
\caption{
Illustration of the discovery of the solar CH/CN anti-correlation
from space- and ground-based observations.
Panel~a displays the spots (left $Y$-axis)
and faculae (right $Y$-axis) fractional filling factors as
a function of time during the Sun's $23^{\mathrm{rd}}$ magnetic cycle.
Panels~b and  c display the corresponding 
mirror-like variability in the CN and CH indices.
Panel~d displays the resultant CH/CN anti-correlation
as detected in space-based observations.
Panel~e corroborates this space-based solar anti-correlation
with ground-based data.
The coordinate spans differ between panels~d and  e 
  due to inherent differences in the spectral continuum between
  ground- and space-based datasets.
}
\label{f_1_chcn_anticor}
\end{figure}

Utilising     the     disc-integrated     SATIRE-S     solar     data,
Fig.~\ref{f_1_chcn_anticor} readily reveals the anticipated empirical
discovery of a CH and CN anti-correlation in the Sun.
To help  the reader appreciate  the significant difference  in spatial
coverage between dark spots and bright faculae, panel~a displays
their  respective  fractional  filling  factors  (spots  on  the  left
$Y$-axis, faculae on the right) as a function of time during the Sun's
$23^{\mathrm{rd}}$ magnetic cycle.
Panels~b and  c of Fig.~\ref{f_1_chcn_anticor} display the
corresponding continuous measurements  of the solar CN  and CH indices
over the same period.
Our full-disc  SATIRE-S reconstruction of the  Sun's molecular indices
throughout  magnetic cycle  \#23  reveals  remarkably mirrored  trends
(compare panels~b and  c).
The   selected  data   points---highlighted   in   blue,  green,   and
red---clearly illustrate that during  periods of low magnetic activity
($\sim$2009),  the  integrated  solar  spectrum  appears  CH-rich  and
CN-poor. Conversely, this behaviour  distinctly reverses to CH-poor and
CN-rich during high magnetic activity ($\sim$2002).
Thus,  at peak  magnetic  activity, the  blue  circles characterise  a
photospheric  environment   where  CH  molecules   are  preferentially
depleted, while CN molecule formation is enhanced.

Directly plotting these indices  against one another, panel~d
reveals a clear CH  and CN anti-correlation---a standard observational
proxy for the C and N anti-correlation in stellar populations.
This demonstrates how even  a relatively inactive, isolated star
 such as  our  Sun  can  naturally  produce  an  apparent  C-poor,  N-rich
spectroscopic signature when  experiencing stronger magnetic activity,
with this signature reverting  under relatively weaker, more quiescent
conditions.

%
Remarkably, this empirical finding is not limited to space-based data or 
restricted to a single magnetic cycle. While the SATIRE-S reconstruction 
spans solar cycle~\#23, ground-based HARPS-N observations gathered at 
the Telescopio Nazionale Galileo (TNG) cover cycle~\#24; despite these 
distinct, non-overlapping epochs, both datasets consistently confirm the 
same fundamental trend (panel~e).
The bottom panel of Fig.~\ref{f_1_chcn_anticor} summarises our
reduction of the HARPS-N solar spectra, displaying the inferred CH and
CN indices for each individual spectrum in the two-epoch
dataset.
Notably, these data illustrate that when the Sun is near its
maximum activity—a phase when facular surface coverage is 
${\sim}$$10$ times greater than that of sunspots—the resulting trend
yields a relatively higher CN index and a lower CH index.
While   the   ground-based  observations   (panel~e)
  robustly reproduce the cycle-driven anti-correlation revealed by the
  ultra-precise space-based data  (panel~d), the two datasets
  naturally  occupy distinct  coordinate spans.
  An order-of-magnitude difference in the  dynamic range of the CN4142
  index  arises despite  the  application  of identical  normalisation
  windows to both datasets.
  It is fundamentally  driven by inherent differences  in the spectral
  continuum definition  and instrumental response profiles  of ground-
  versus space-based configurations.
  The  difference is  particularly  pronounced  across the  relatively
  broad  $88$\,\AA\  CN4142 index  window,  unlike  the much  narrower
  $15$\,\AA\  CH4300  bandpass. The  wider  CN4142  window is  highly
  susceptible to substantial stray light contamination and atmospheric
  scattering as both  the true solar flux  and instrumental throughput
  decrease sharply below $4300$\,\AA.

An attentive reader might note that the nominal activity maximum of solar 
cycle~\#$24$ occurred in $2013$, whereas the earliest high-activity HARPS-N 
dataset is from July $2015$.
Crucially, despite capturing the Sun approximately two years after its
peak, the available data still allow us to recover a clean, well-separated
CH/CN anti-correlation.
However, a closer examination of detailed magnetic field maps
\citep{yoshida2023} reveals that even a modest time difference
of $\sim$$2$\text{--}$3$\,years around the nominal cycle
maximum can correspond to a dramatic shift in the Sun's
average magnetic field configuration.
Specifically, the solar field 
evolved from a completely absent polar contribution in $2013$ to a 
well-defined open solar flux by $2015$, prominently extending from the Sun's 
polar regions. This observation perfectly aligns with the analysis by 
\citet{stansby2021} of photospheric magnetic field maps spanning four 
solar cycles. They concluded that at any given time during solar maxima, 
the fractional contribution of active regions and faculae to the open 
solar flux can vary substantially, ranging from $30$$\%$ to $80$$\%$.
Consequently, while our July 2015 observations robustly confirm the 
existence of the CH/CN anti-correlation, the inferred amplitude likely 
represents a conservative lower bound on the full peak-to-minimum 
spectroscopic variation.

Having  established the  unambiguous existence  of a  solar CH  and CN
anti-correlation,  in Appendix\,\ref{s_solar_fundamentals}  we briefly
examine the fundamental properties  of the primary magnetic components
(faculae  and sunspots)  that  drive this  molecular modulation.  This
overview provides  the physical baseline for  interpreting these solar
spectral variations.

  \section{The potential relevance of  faculae to the
    globular cluster  conundrum}
\label{s_faculae_vs_globulars}

Given  that    solar  brightness  variations   are  fundamentally
  faculae-dominated   \citep{shapiro2016,   cretignier2024},  it   is
reasonable  to expect  that this  magnetic behaviour  extends to  other
low-mass stars, including those within GCs.
In  turn,   this  would   imply  that  small-scale   surface  magnetic
fields---manifesting   as  faculae---may   play  a   significant,  yet
previously overlooked, role in the MP conundrum.

\subsection{The facular link to the MP conundrum}
\label{s_mps_faculae_link}

\begin{figure}
\centering
\includegraphics[width=0.95\hsize]{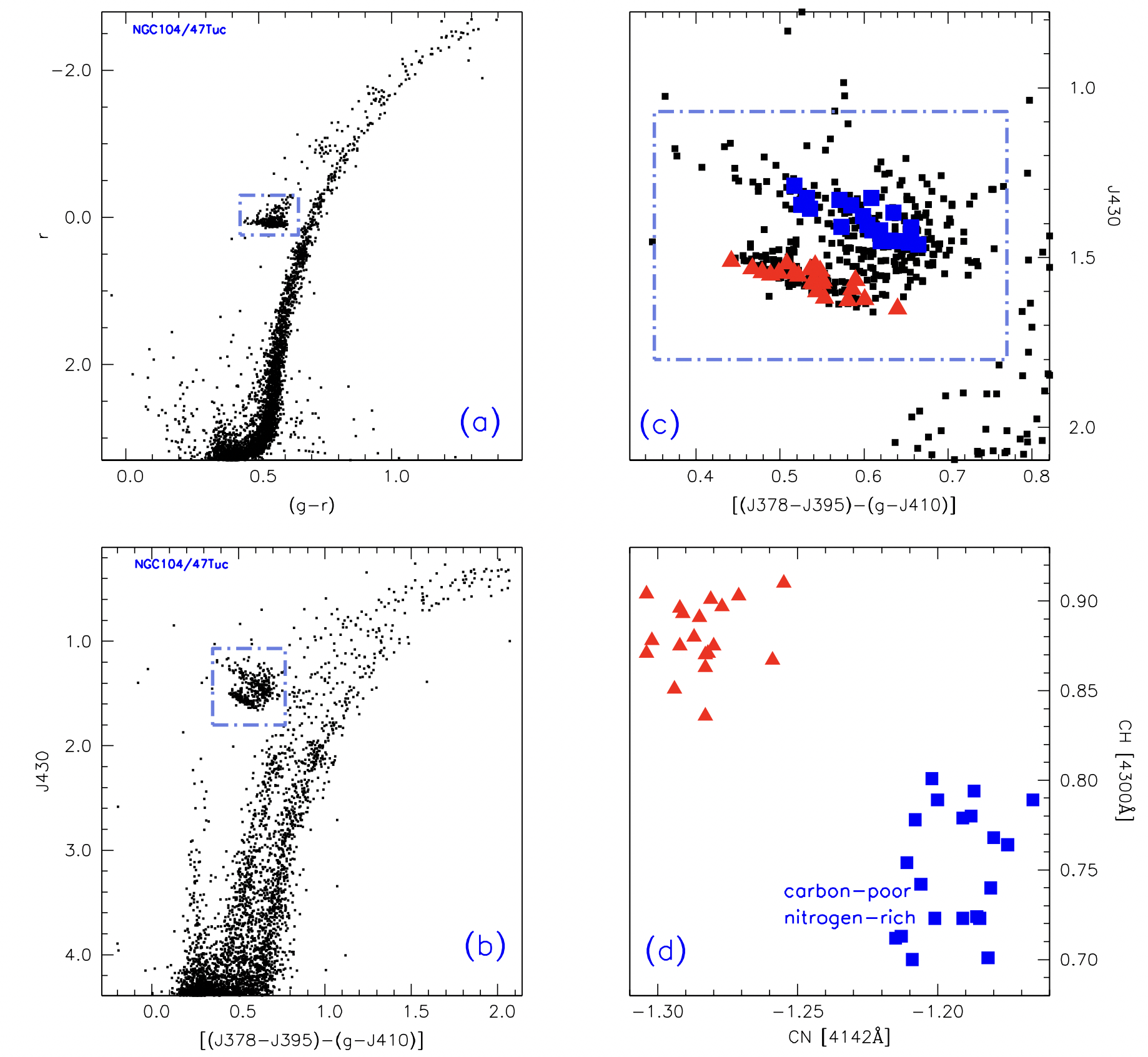} 
\caption{
  Illustration of the MP conundrum in GCs. 
  Panel~a displays the classical colour-magnitude diagram of
  NGC104, where a single red giant branch and a
  single red horizontal branch sequence are visible.
  Panel~b re-plots the  same stars using S-PLUS
  narrow-band filters, revealing two distinct sub-populations.
Panel~c provides  a zoomed-in view of the  newly uncovered double-deck
red-HB    morphology,    highlighting    the    \citep{villanova2025x}
targets.
Panel~d displays the  CH and CN molecular  band distribution
for these  stars.
Combined, panels~c and   d illustrate that 2P  stars (blue
symbols)  are relatively  more  CH-deficient,  CN-enriched, and,  most
importantly,  G band  continuum-enhanced  compared to  1P  stars  (red
symbols)—collectively consistent with the hallmark observational traits of
a solar facular signature.
}
\label{f_3_conundrum}
\end{figure}

Figure~\ref{f_3_conundrum} illustrates our proposed parallel between 
the properties of solar faculae and the signatures of 2P stars, while 
highlighting the MP conundrum in bright globular cluster stars.
We  specifically   focus  on   the  more   luminous  regions   of  the
colour-magnitude  diagram to  highlight  a  defining 2P  characteristic
predicted by \citet{sbordone2011}: an  enhanced G band luminosity that
persists across both evolved and unevolved evolutionary stages.
In  this  regime,  2P  red-HB  stars  exhibit  an  unambiguous  G band
luminosity enhancement; a distinct  photometric jump.  Crucially, such
an  enhanced  G band  luminosity  is also  the  hallmark  of  magnetic
faculae.

Specifically,  Fig.~\ref{f_3_conundrum}  captures the  paradigm  shift
from   the  classical   view  of   globular  clusters   as  chemically
homogeneous,  coeval `simple  stellar  populations'  to the  complex
reality  of  multiple  sequences unravelled  by  modern  high-precision
photometry.
This transition is demonstrated by the contrast between panels~a and b
of  Fig.~\ref{f_3_conundrum}.   While  standard  visual  $g$  and  $r$
filters yield a traditional, unimodal  red giant branch (panel~a), the
combined near-ultraviolet  and optical narrow-band T80  filters unveil
the dual nature of these very same stars (panel~b).
This bifurcation is particularly  evident in the red-horizontal branch
(red-HB) stars, as shown in the enlarged view of panel~c.
The  distribution  of  red-HBs   reveals  a  distinct  `double-deck'
morphology: the  2P population  (blue symbols) appears  roughly $0.15$
magnitudes   brighter  in  the  G band/J430  filter  than  their
canonical 1P counterparts (red  symbols).
Although both groups occupy  a similar `pseudo-colour' region—defined
by                the                 four-filter                index
[($J378$$-$$J395$)$-$($g$$-$$J410$)]        in       panels~b 
and~c—the  2P  red-HB stars  are  found  to be  approximately
$100$\,K hotter than the 1P population (\citealt{villanova2025x}).
Panel~d  displays the  corresponding  CH/CN  anti-correlation for  the
red-HB stars highlighted in panel~c.
  In this  chemical space, the  2P (CH-poor/CN-rich) stars  cluster in
  the   lower-right  (blue   symbols),   while   their  canonical   1P
  (CH-rich/CN-poor) counterparts occupy the upper-left (red symbols).

In  summary,  the  defining  observational  traits  of  solar  facular
regions—namely, intrinsic  CH deficiency, CN enrichment,  and enhanced
G band   continuum    emission—share   striking   qualitative
  similarities   with   the   distinctive   spectroscopic   and
  photometric  properties  observed  in   2P  stars  across  GCs.
  Observational precedent already exists for this parallel. Indeed, compelling 
empirical evidence anticipating a correlation between surface 
inhomogeneities and magnetism in GC 2P stars has been provided by 
\citet{momany2020}, who, although tracking an exotic evolutionary phase 
via extreme horizontal branch stars, discovered widespread 
rotational variables inferred to host giant, bright, and hot magnetic 
spots that perfectly mirror the localised thermodynamic properties of 
solar faculae.

\subsection{Scaling  the solar  imprint:  Simulations  of an  enhanced
  magnetic state}
\label{s_simul}

It may be argued  that the variations in the solar  CH and CN indices,
and  the   amplitude  of the  resulting  CH/CN  anti-correlation
(panels~d  and e  in  Fig.~\ref{f_1_chcn_anticor}),
represent  only a   miniature  reflection of  the significantly  larger
spectroscopic spreads observed in  globular clusters. This observation
is undeniably accurate.

From  this   perspective,  the  parallelism  we   propose—between  the
properties of  solar faculae and  the signatures of 2P  stars—could be
viewed   as  a   compelling   but   ultimately  coincidental   scaling
discrepancy.
To explore  the validity of this  link, we had to address  a fundamental
question,  namely  whether  the CH/CN  anti-correlation  persists  and
scales accordingly if the magnetic `volume' of the Sun is turned up.

To investigate this, we turn to numerical simulations, which currently
represent the only viable avenue for such an exploration.
Our  approach is  grounded in  the  observation that  the Sun  remains
remarkably   quiescent  compared   to   other  solar-type   analogues,
notwithstanding  the   energetic—and  occasionally  explosive—magnetic
phenomena observed at its surface.
Indeed, the  Sun would  likely not  even be  flagged as  `active' by
current Kepler detection algorithms \citep{reinhold2020}.

We therefore moved  into uncharted terrain by  performing a first-order
diagnostic simulation.   In this framework, we  systematically enhanced 
the  average   surface  magnetic   field  of  a   solar-analogue  model,
artificially  increasing  the  fractional   coverage  of  faculae  and
sunspots.   Our goal  was not  to produce  a high-fidelity  model of  a
globular  cluster  star, but  rather  to  investigate the  first-order
spectroscopic  imprints that  a higher  magnetic filling  factor would
impose upon the established solar CH/CN anti-correlation.
The    details     of    the     simulation    are     presented    in
 Appendix\,\ref{s_details_simul}.

\begin{figure}
\centering
\includegraphics[width=0.89\hsize]{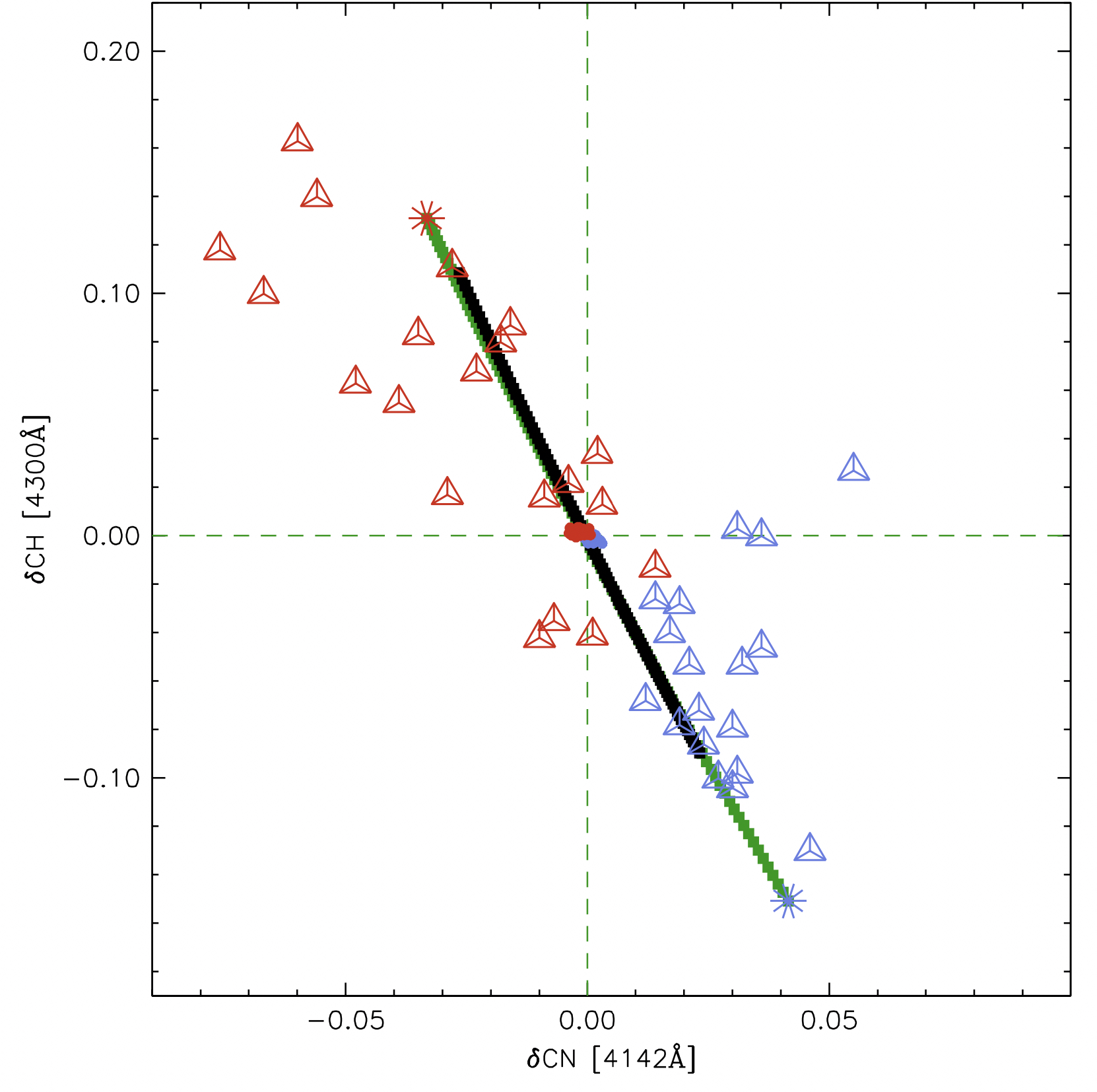} 
\caption{
Illustration of how simulating a magnetically enhanced solar-model amplifies
the Sun's CH/CN anti-correlation.
Filled red and blue circles clustering at ($0$,$0$) plot the solar
ground-based CH/CN distribution (panel~e of Fig.~\ref{f_1_chcn_anticor}),
after subtraction of its centroid.
Black symbols display the $\delta$CH/$\delta$CN distribution from a
solar-model simulation that includes only the magnetically enhanced
contributions of faculae. In contrast, green symbols represent a similar
simulation that incorporates the combined magnetically enhanced
contributions of both faculae and sunspots.
Blue and red asterisk symbols denote the opposite extremes
of the $\langle$$B$$\rangle$$=$$3000$ and $0$\,Gauss simulations,
respectively.
Double-triangle symbols display empirical $\delta$CH/$\delta$CN
measurements of dwarf stars in NGC104 \citep[cf. Fig.~7\,in][]{pancino2010}.
Blue and red symbols correspond to 2P and 1P stars, respectively.
}
\label{f_4_simul_sun}
\end{figure}

 To   ease    a   meaningful    comparison   across    the   disparate
 datasets---specifically   the   solar   HARPS-N   observations,   the
 magnetohydrodynamic (MHD) simulations, and  the FORS2 observations of
 NGC104  (\citealt{pancino2010})---we  adopted  differential  indices,
 $\delta$CH and $\delta$CN.
By centring each distribution on its respective centroid, we mitigate
systematic offsets arising from  differences in spectral normalisation
and instrumental  response. This standard procedure  \citep[cf. Fig.~7
in][]{pancino2010} enables  a direct cross-comparison of  the relative
amplitudes and  slopes of  the CH/CN anti-correlation,  independent of
the underlying spectral properties of the individual datasets.

Figure~\ref{f_4_simul_sun} presents  the results of  two complementary
simulations for  a magnetically enhanced `active'  Sun: one accounting
for faculae alone and  another incorporating the combined contribution
of faculae and sunspots.
Each data  point corresponds  to a  specific magnetic  field strength,
$B$.     Notably,      the     points     at      $B$$=$$0$\,G     and
$B$$=$$3000$\,G   represent    theoretical,   physically   unrealistic
extremes:  a star  entirely devoid  of  magnetic features  and a  star
completely  covered by  spots  and faculae,  respectively.  These  two
endpoints  define  the  maximum  theoretical amplitude  of  the  CH/CN
variation within the modelled green distribution.

The green  distribution in Fig.~\ref{f_4_simul_sun}  demonstrates that
incorporating both  sunspots and faculae does  not significantly alter
the slope established by the faculae-only model (black points).
Sensitivity  tests  exploring  a range  of  filling  factors—extending
beyond                            the                            equal
$50$$\%$:$50$$\%$     faculae-to-spot     ratio     illustrated     in
Fig.~\ref{f_4_simul_sun}—confirm that while  the amplitude of
the CH/CN  anti-correlation scales  with total magnetic  coverage, the
underlying slope remains remarkably robust.

Regardless  of  the weight  assigned  to  a  facular  role in  the  MP
conundrum, the comparison between the cycle-long solar HARPS-N spectra
and      our      magnetically      enhanced      MHD      simulations
(Fig.~\ref{f_4_simul_sun})  yields a  fundamental insight:  increasing
stellar magnetic  activity directly drives a  significant expansion in
the CH/CN anti-correlation amplitude.
Notably, from  a quantitative  standpoint, our  combined model---which
accounts   for    the   simultaneous    presence   of    faculae   and
sunspots---yields a  spectroscopic spread  that is  broadly consistent
with  the empirical  FORS2  observations\footnote{We  are grateful  to
  E. Pancino for kindly providing  us with her complete $\delta$CH and
  $\delta$CN   NGC104   catalogue.}   of   dwarf   stars   in   NGC104
\citep{pancino2010}.

While we acknowledge the  inherent simplifications of this first-order
diagnostic approach, the alignment  between these idealised models and
cluster data provides a notable point of consistency.
Indeed, the fact that a framework rooted entirely in solar physics, 
scaled to account for higher activity levels \citep{reinhold2020}, 
can reproduce the generalised slope and scale of the variations in 
NGC104 underscores the physical viability of this hypothesis.
Ultimately,   Fig.~\ref{f_4_simul_sun}   demonstrates   that   surface
magnetism—encompassing both small- and large-scale features—provides a
viable, self-consistent  physical framework  capable  of driving
the CH/CN anti-correlation in globular clusters.

\section{Final remarks and future perspectives}
\label{s_conclusion}

In  summary,  the  solar   surface  hosts  three distinct  magnetic  regimes
characterised  by increasing  field strengths  and decreasing  filling
factors:                           a                          `weak'
${\sim}170$\,G       photospheric       background,       `moderate' 
${\sim}1000$\,G      widespread      faculae,      and      `strong' 
${\sim}3000$\,G localised sunspots.
This raises a fundamental question regarding why such pervasive surface 
fields, irrespective of their individual filling factors, do not elevate the 
Sun's global magnetic field beyond its modest measured value of 
${\sim}2$\,G (\citealt{plachinda2011}).
The  answer  lies  in  the  intrinsic  inhomogeneity  of  the  surface
plasma.  When high-strength  magnetic elements  are intermingled  with
non-magnetic regions, polarisation-based diagnostics—which rely on net
vector signals—experience significant  flux cancellation.
This  results in  a  deceptively low  global  magnetic signature  that
belies the true  physical strength of the surface  features.  As noted
by \citet{sanchez2000}, this realisation  implies that  `enhanced
  continuum emission may  be a more reliable tracer  of magnetism than
  line polarisation.'

  It is precisely this  recognition---that enhanced continuum emission
  serves as  a more  robust tracer of  magnetism---that allowed  us to
  identify the striking parallel with the G band continuum enhancement
  predicted by \citet[cf. their  Figs.~4 and 5]{sbordone2011} for both
  evolved and unevolved 2P stars.
The immediate  visual evidence of an  enhanced G band continuum
in  the  red-HB 2P  population  of  NGC104 (cf.  panel~c  of
  Fig.~\ref{f_3_conundrum})  ultimately  prompted  us  to  explore  a
potential magnetic origin for the broader MP phenomenon.

Following this line of reasoning,  we here demonstrate that the active
Sun,  as  a  single  star,   exhibits  a  measurably  CH-depleted  and
CN-enriched spectroscopic signature relative to its quiescent state.
This  solar  signature  is   intrinsically  linked  to  a  photometric
manifestation: an enhanced G band continuum emission.
These  faculae-dominated  solar  signatures  closely  align  with  the
primary  characteristics of  2P stars,  suggesting  a potential
role of  surface magnetism  in the MP  phenomenon.
This connection is further  reinforced by our first-order simulations,
which  indicate  that  solar-type  CH and  CN  variations   scale
  substantially under higher magnetic filling factors.
Notably,  while  not  an  explicit objective  of  this  modelling,  the
simulations appear to broadly match the spectroscopic spreads observed
between 1P and 2P populations in globular clusters.

Overall, the results presented in this paper introduce a new dimension
to `surface inhomogeneity'  scenarios---frameworks that were largely
dismissed decades ago when chemical anomalies were found to persist in
both   evolved  giants   and   un-evolved  dwarfs   \citep{briley2004a,
  briley2004b}.
Unlike those historical models, however, which relied on stochastic external
contamination (such as the accretion of interstellar material or
planetary engulfment; cf. \citealt{liu2024, carlos2025}), we propose an
intrinsic, self-regulating, and astrophysically grounded mechanism.

Given           that            all           low-mass           stars
(M$\lesssim$$1.1$\,M$_\odot$)     possess     convective     envelopes
\citep{renzini1988,    chiosi1992}---the   fundamental    engine   for
dynamo-driven surface magnetism \citep{brun2017, charbonneau2020}---it
is logically consistent  to expect that the  solar magnetic phenomenon
extends    to    all    structurally   similar    stellar    analogues
(M$\lesssim$$0.85$\,M$_\odot$) within GCs.
This realisation, coupled with the well-established connection between
convective-envelope structure, rotation, and  the absence of anomalies
in                clusters                 younger                than
$\sim$$2$\,Gyr, further strengthens the  case for surface magnetism as
a viable mechanism.
Indeed,  as  outlined   in  Appendix\,\ref{s_young_clusters},  surface
magnetism can provide a unified account capable of spanning the entire
evolutionary timeline across which the MP phenomenon is detected.

Crucially, while  the solar cycle  serves as a physical  laboratory to
validate the underlying physics of the  parallel drawn with GCs, it is
not the primary driver of the MP phenomenon itself.
The observed anti-correlations in GCs are not direct manifestations of
solar-like cycles; indeed, during its cycle, a star may `wobble' along
its average  anti-correlation sequence, but  such an effect  is likely
too minor to be detected.
In fact, an ongoing comparison between  two WFI/2.2m wide-field
NGC104          $BVI$          datasets          separated          by
$\sim$20 years appears to confirm  the absence of P1/P2 average-colours
cross-population migration.
This finding  is corroborated by  the overall luminosity  stability of
HD166620, the  first definitive  extra-solar Maunder  minimum analogue
(see Appendix\,\ref{s_magnetic_cycles}).
Instead, the emerging  role of surface magnetism is  predicated on the
persistent,     intrinsic     magnetic     state     of     individual
stars---distinguishing      relatively       active      from
relatively                  quiescent                 members.
This interpretation thus relies on surface brightness contrasts rather
than   absolute  global   magnetic   field  strengths.   As  the   Sun
demonstrates,  global signatures  can remain  deceptively modest  even
while  intense,  intermingled  magnetic  fields  generate  distinctive
photometric and spectroscopic signatures.

Recognising  the  need for  a  distinct  nomenclature to  separate  an
intrinsic,   complex   magnetic   process   from   extrinsic   surface
contamination theories, we propose the \textsc{MAGISTER\,FAS} acronym:
Magnetic Activity  Generating Inhomogeneous Surfaces Tied  to Enhanced
Relative Faculae And Spots.
Should  the  \textsc{MAGISTER  FAS}  framework be  admitted  into  the
broader  discourse  on  the   MP  phenomenon,  the  spectroscopic  and
photometric  anomalies  defining  these populations  would  no
  longer   necessitate   an   interpretation  based   on   successive
star-formation episodes. Instead, they would emerge as the predictable
manifestation of star-to-star variations in intrinsic surface magnetic
activity levels.

In its current nascent state, the \textsc{MAGISTER FAS} framework 
remains a considerable distance from accounting for the full complexity 
of the MP phenomenon. Specifically, we currently lack the empirical 
constraints required to properly address other key signatures 
(e.g. the Na/O and Mg/Al anti-correlations) and the detailed physics 
of how surface magnetic imprints are maintained across evolutionary 
transitions.
However, while this work focuses primarily on molecular tracers, 
we note that the underlying thermodynamic mechanisms of the 
\textsc{MAGISTER FAS} framework are expected to also affect key 
atomic abundance indicators (\ion{O}{i}, \ion{Na}{i}, \ion{Mg}{i}, 
and \ion{Al}{i}).
  To explore  how magnetically induced atmospheric  stratification can
  lead   conventional   one-dimensional   (1D)   local   thermodynamic
  equilibrium (LTE) radiative transfer analyses to misinterpret subtle
  profile  variations   as  spurious  chemical  spreads---and   why  a
  transition  towards  three-dimensional (3D)  non-local  thermodynamic
  equilibrium   (non-LTE)   magnetic  radiative   transfer   modelling
  represents an  important next  step for the  cluster community---the
  reader   is   directed   to    the   comprehensive   discussion   in
  Appendix~\ref{s_atomic_variations}.

Consequently, we present this framework not as a definitive resolution, but as 
a physically motivated, solar-anchored proposal that any future unified theory 
of the MP phenomenon must necessarily account for. Even if regarded as a 
second-order effect, these magnetic signatures  are an unavoidable 
consequence of the convective-envelope architecture inherent to all low-mass 
stars in globular clusters.
Given this physical mandate, we draw attention to independent lines of 
evidence where solar-like magnetism likely plays a role---at least 
qualitatively---in defining MP properties:

\begin{enumerate}

\item[(i)] The faculae-dominated signatures  identified in the Sun are
  theoretically predicted  to scale with spectral  type, reaching high
  contrast towards the  cool end of the G-type sequence and peaking in
  early K-type  dwarf stars,  before declining to  a minimum  near the
  K/M-dwarf       boundary       \citep{steiner2014,       norris2023,
    kostogryz2026, godoyrivera2026}.
  This predicted K-type facular efficiency peak aligns remarkably well
  with  the stellar  regime where  the maximum  photometric separation
  between 1P  and 2P  sequences is  empirically observed  (roughly two
  magnitudes below the main-sequence turnoff, near $5300$\,K).

\item[(ii)]  MHD  simulations  indicate a  distinctive  transition  in
  facular      properties      within     the      M-dwarf      regime
  \citep{shapiro2026}.  This  behaviour provides  a  qualitative 
  physical basis  to suggest a  magnetic origin for both  the observed
  main-sequence  splitting  and   the  intriguing  colour-reversal
  between  1P   and  2P   populations  in   M-type  stars   (see
    Appendix\,\ref{s_mtype}).

\item[(iii)]  In Appendix\,\ref{s_7indici},  we  show that  the
  solar CH and CN bands, as  well as the OH and NH features, are
  all          significantly         sensitive          to         the
  $\pm$$1000$\,K   temperature   fluctuations   induced   by   surface
  magnetism.
  Consequently, the characteristic  splitting of photometric sequences
  in  globular  clusters---consensually   attributed  to  the
    differential response  of the molecular bands  within the employed
    filters---appears   interpretable  as   a  natural   manifestation  of
  differential magnetic activity.

\end{enumerate}

Despite decades of scrutiny, the  Sun is still an enigmatic benchmark;
notably, the fundamental mechanisms  governing its internal dynamo and
cycle-to-cycle   variability   are    still   not   fully   understood
(\citealt{charbonneau2020}).
Similarly, questions about long-term variability and the nature of grand 
magnetic minima---dormant states into which the Sun regularly transitions, 
spending nearly $17\%$ of its lifetime \citep{usoskin2017}---remain outstanding.
Such persistent uncertainties further underscore the challenges inherent in 
extrapolating solar magnetism to the diverse temperatures, ages, and 
metallicities of stars in globular clusters.
This is compounded by the fundamental difficulty of transposing the magnetic 
behaviour of an isolated field star such as the Sun onto the dense environments of 
clusters—where, for instance, a volume equivalent to the void space separating 
the Sun from Proxima Centauri would encompass several hundred thousand stars in 
the core of NGC104.
In  such  extreme  conditions,   members  are  inevitably  subject  to
continuous, prolonged,  and stochastic  dynamical interactions  over a
Hubble  time---environmental complexities  that, while  challenging to
model, must ultimately be accounted for.
Within our \textsc{MAGISTER FAS} framework, these persistent 
encounters would provide a highly efficient physical engine for angular 
momentum injection, which ultimately enhances the surface magnetic 
activity predicted to exist within globular cluster environments.
Interestingly, direct empirical evidence of this hypothesised 
environment-driven spin-up process in GCs is provided by the 
exceptionally short rotation periods detected among the BY\,Dra 
variables in NGC104 (cf.\ Appendix~\ref{s_dynamics}).

Crucially,    reported    detections     of    2P-like,    superficial
$\mathrm{Na}/\mathrm{O}$     signatures    within     open    clusters
\citep{pancino2018}  and wide-field  Galactic surveys  find a  natural
physical explanation within our  proposed scenario.  Indeed, under the
\textsc{MAGISTER  FAS} framework,  multiple-population signatures  are
fundamentally  governed  by the  internal  activation  of an  enhanced
stellar rotational state.
While  the  dense  core  dynamics  of  globular  clusters  permanently
maintain the  necessary rapid  rotation across cosmic  timescales, the
lower-density regimes of open clusters  and the Galactic field capture
these  identical  chemical  anomalies only  as  fleeting  evolutionary
snapshots  or  rare,  stochastically generated  products  where  rapid
stellar  rotation  is briefly  achieved  in  single configurations  or
tidally preserved through close binary interactions. Fundamentally, it
is the  persistent environmental injection of  angular momentum unique
to  GC cores  that  effectively  overrides standard  gyrochronological
spin-down,  transforming what  would otherwise  be a  brief, transient
stellar phase into a long-lived, structurally fixed configuration.
In  this  way,  \textsc{MAGISTER  FAS}   provides  a  sound  basis  to
hypothesise  an  empirical  globular-to-open-cluster-to-field  2P-like
link that elegantly connects the macro-environmental dynamics of these
diverse systems back to  the fundamental core-envelope stellar physics
detailed in Appendix~\ref{s_young_clusters}.

At present, only the initial implications of the \textsc{MAGISTER FAS}
framework  have  been  explored—a   framework  that  is  fundamentally
ubiquitous among the  low-mass stars of globular  clusters. While this
work serves as a proof-of-concept, it ultimately calls for a necessary
synthesis between the  solar and stellar communities  to determine the
impact  of   this  magnetic  perspective  on   the  broader  discourse
surrounding the MP conundrum.

\begin{acknowledgements}
  We thank  the referee,  Dr. Nate  Bastian, whose  insightful
  comments and  valuable suggestions  were instrumental  in sharpening
  our arguments and significantly improving  the final version of this
  paper.
  We  warmly remember  Paola  Marigo,  whose thoughtful  conversations
  accompanied the  early stages of  this work.  Her passing  is deeply
  mourned.
   We gratefully acknowledge  E. Carretta, S. Cassisi,  and M. Salaris
   for generous  and enriching discussions.
   We  also thank  A.  Shapiro  and N.   Kostogryz  for their  crucial
   guidance and  contributions to  the solar  simulations, as  well as
   K.  Sowmya for  her  assistance in  measuring  the solar  molecular
   indices.
%
   L.M.  gratefully acknowledges  support  from ANID-FONDECYT  Regular
   Project n. 1251809.
   S.V.  gratefully  acknowledges  the support  provided  by  Fondecyt
   Regular n. 1220264 and by the ANID BASAL project FB210003.
   L.G., S.Z. and Y.M.  acknowledge support from projects `SPICE4LSST
   –  Adding  incompleteness and  crowding  errors  to the  LSST  Data
   Releases' (INAF Large Grant  2024) and `Population synthesis with
   rotating stars: a necessary upgrade' (INAF Theory Grant 2022).
   N.R.L. thanks  the Brazilian agencies  CAPES, CNPq and  FAPEMIG for
   financial support,   and thanks  Dr. Francesca  D'Antona (INAF-OAR,
   Italy) and  Italo Mazzitelli  (in memoriam)  for granting  her full
   access to  the ATON  evolutionary code.
   %
   G.C. acknowledges  financial support from the  European Union--Next
   Generation  EU,  Mission  4,  Component  2,  CUP:  C93C24004920006,
   project 'FIRES'.
   F.R.H.  acknowledges support  from FAPESP  grants 2018/21661-9  and
   2021/11345-5 and the T80-South technical team for the support with the observations.
   F.A.-F. acknowledges support from  FAPESP grant 2024/00822-5, from the Brazilian Ministry of Science, Technology and Innovation (MCTI) and the Brazilian Space Agency (AEB), which supported the present work under the PO 20VB.0009.
   Language refinement was supported by Google Gemini.
\end{acknowledgements}


%
\bibliography{aabib} 
\bibliographystyle{aa}

%
\begin{appendix}

\section{The solar magnetic baseline}
\label{s_solar_fundamentals}

\subsection{The differences between faculae and spots}
\label{s_faculae_vs_spots}
Sunspots are the most conspicuous manifestations of the Sun's 
 large-scale magnetic fields.
Compared    to    the   photosphere's    $170$\,G    baseline
  \citep{zeuner2024},      sunspots    harbour   intense   fields   of
${\sim}2000{\div}4000$\,G that depress the local temperature by nearly
${\sim}1000$\,K  compared to  the  quiet Sun,  making them  
  darker and easily discernible.
Over two  centuries of systematic  sunspot monitoring have  revealed a
quasi-periodic  ${\sim}11$-year  cycle  that governs  their  rise  and
decline  in number\footnote{It  is important  to note  that the  solar
  magnetic  cycle  deviates  significantly from  a  simple  sinusoidal
  pattern.  It is often  marked by double-peaked maxima---arising from
  activity   differences    between   the   northern    and   southern
  hemispheres---and exhibits  an asymmetric rise and  decline relative
  to solar minima \cite{hathaway2010}.}.

Measurements  of  the total  solar  irradiance  (TSI) over  the  Sun's
${\sim}27$-day rotational period prove that the transit of dark,
cool  sunspots   across  the   solar  disc   leads  to   a  measurable
${\sim}0.3\%$ TSI  dimming.
Yet,  on longer  ${\sim}11$-year  timescales,  an intriguing  reversal
emerges: at the peak of magnetic activity, when as many as ${\sim}150$
sunspots may  appear in a  single year, the  Sun's TSI exhibits  a net
${\sim}0.1\%$ brightening.
This raises  a paradox:  how can sunspots--responsible  for short-term
TSI   dimming--coexist  with long-term  TSI   brightening,
especially given their sheer abundance at maximum activity?
The  resolution  lies in  the  contribution  of the  often  overlooked
small-scale magnetic fields, or faculae.

Faculae are the  luminous counterparts of sunspots,  appearing as hot,
bright regions rather than cool, dark depressions.
With                 magnetic                fields                 of
$500{\div}1500$\,G, they are  weaker (\citealt{beck2007}) than sunspots
yet invariably  accompany them, often emerging  earlier and persisting
longer (\citealt{foukal2013}).
Moreover,    as     illustrated    in    the    upper     panels    of
Fig.~\ref{f_2_fac_spo},  faculae  occupy   a  vastly  larger  surface
fraction than that of sunspots.
It  is  this  indispensable   collective  brilliance  of  the  facular
contribution that  offsets the  radiative deficit imposed  by abundant
sunspots,                yielding               the                net
${\sim}0.1\%$   increase    in   TSI--characteristic   of    our   Sun
 (\citealt{ermolli2013,solanki2013,kopp2016}).
 Narrow-band imaging  in the Fraunhofer  ($429.5$$-$$431.5$\,nm) G
 band provides  one of  the most effective  methods for  tracing solar
 faculae.
 This spectral region is dominated  by CH molecular lines, meaning the
 integrated  brightness  is heavily  modulated  by  the abundance  and
 opacity of  these features.   In the context  of the  `bright wall'
 model \citep{spruit1976}, the plasma within magnetic flux
 tubes—the physical  manifestation of faculae—is  significantly hotter
 than the surrounding quiet photosphere.
 Under  thermodynamic  equilibrium,  the  abundance of  CH  is  highly
 sensitive  to  local   temperature  and  density  \citep{sanchez2001,
   steiner2001, langhans2002}. Consequently, the elevated temperatures
 within  facular  regions  trigger  the  thermal  dissociation  of  CH
 molecules.
 This depletion  of CH absorption effectively  `unmasks' the deeper,
 brighter layers, resulting in a strong intensity contrast relative to
 the  cooler, CH-rich  surroundings.
 This    underlying   mechanism    was   robustly    demonstrated   by
 \citet{shelyag2004}, who  traced the evolution  of CH abundance  as a
 function  of magnetic  flux.  Furthermore,  \citet{yeo2013} confirmed
 that  the  facular contribution  to  solar  irradiance variations  is
 driven by the combined effects of continuum excess and these specific
 molecular line changes.

\subsection{Evolution of facular and sunspot surface coverage}
\label{s_coverage_fac_spo}
\begin{figure}
\centering
\includegraphics[width=0.99\hsize]{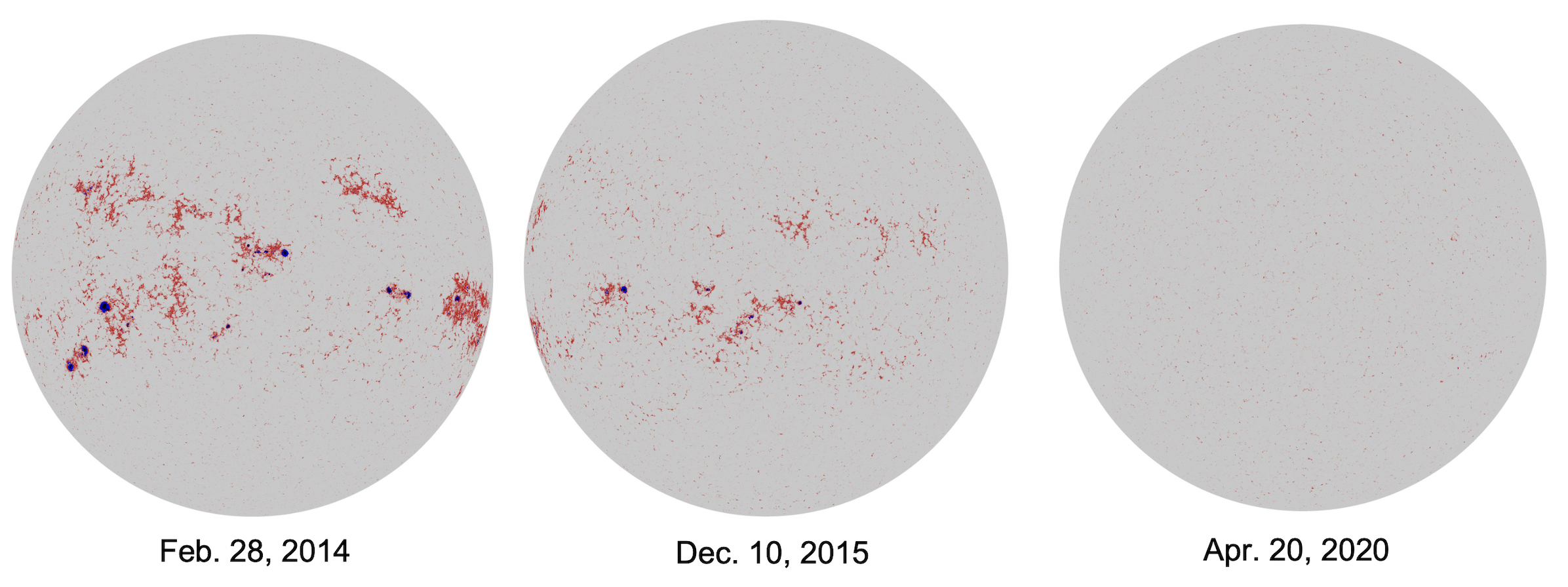}
\includegraphics[width=0.44\hsize]{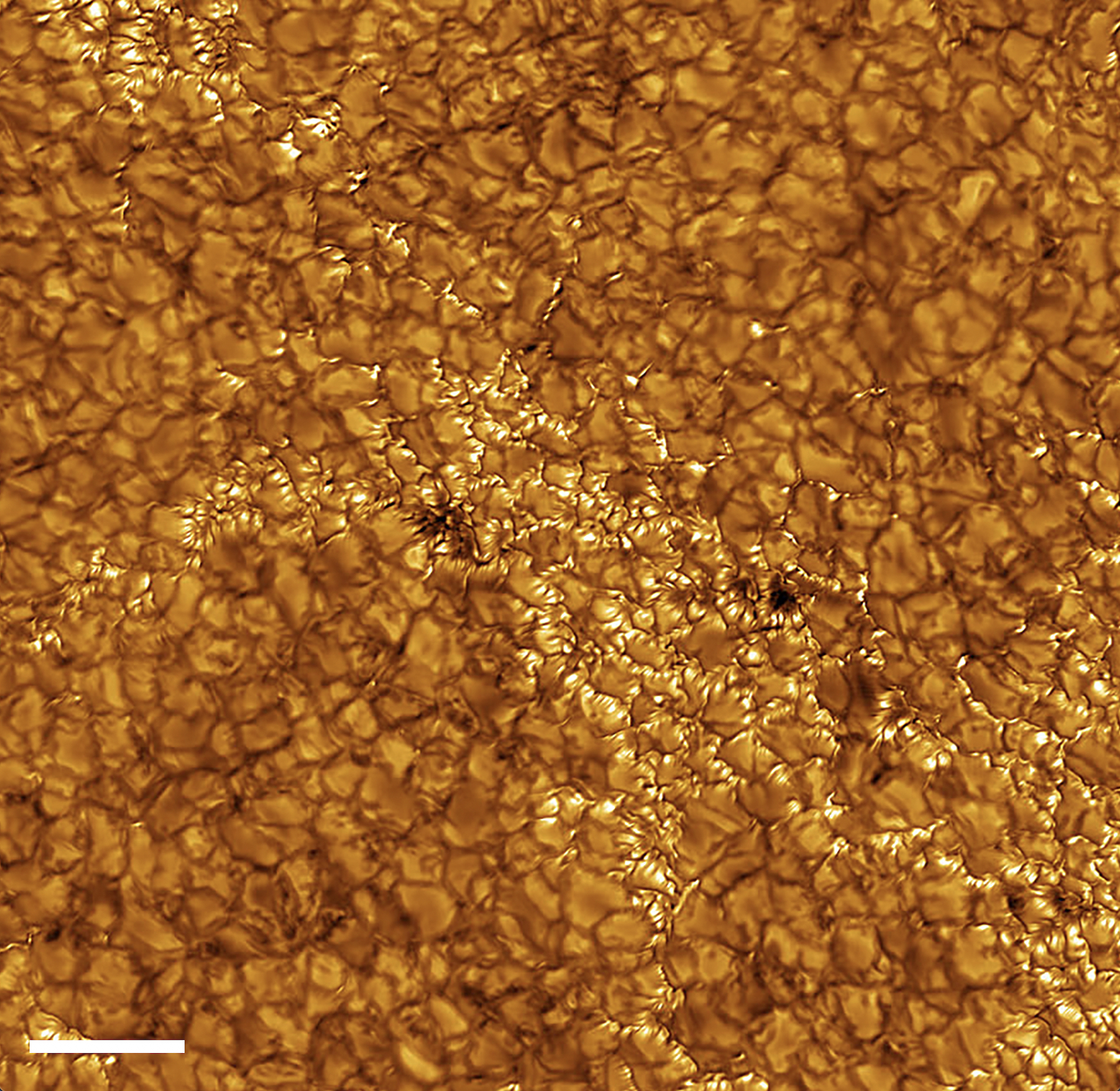} 
\includegraphics[width=0.44\hsize,height=0.155\vsize]{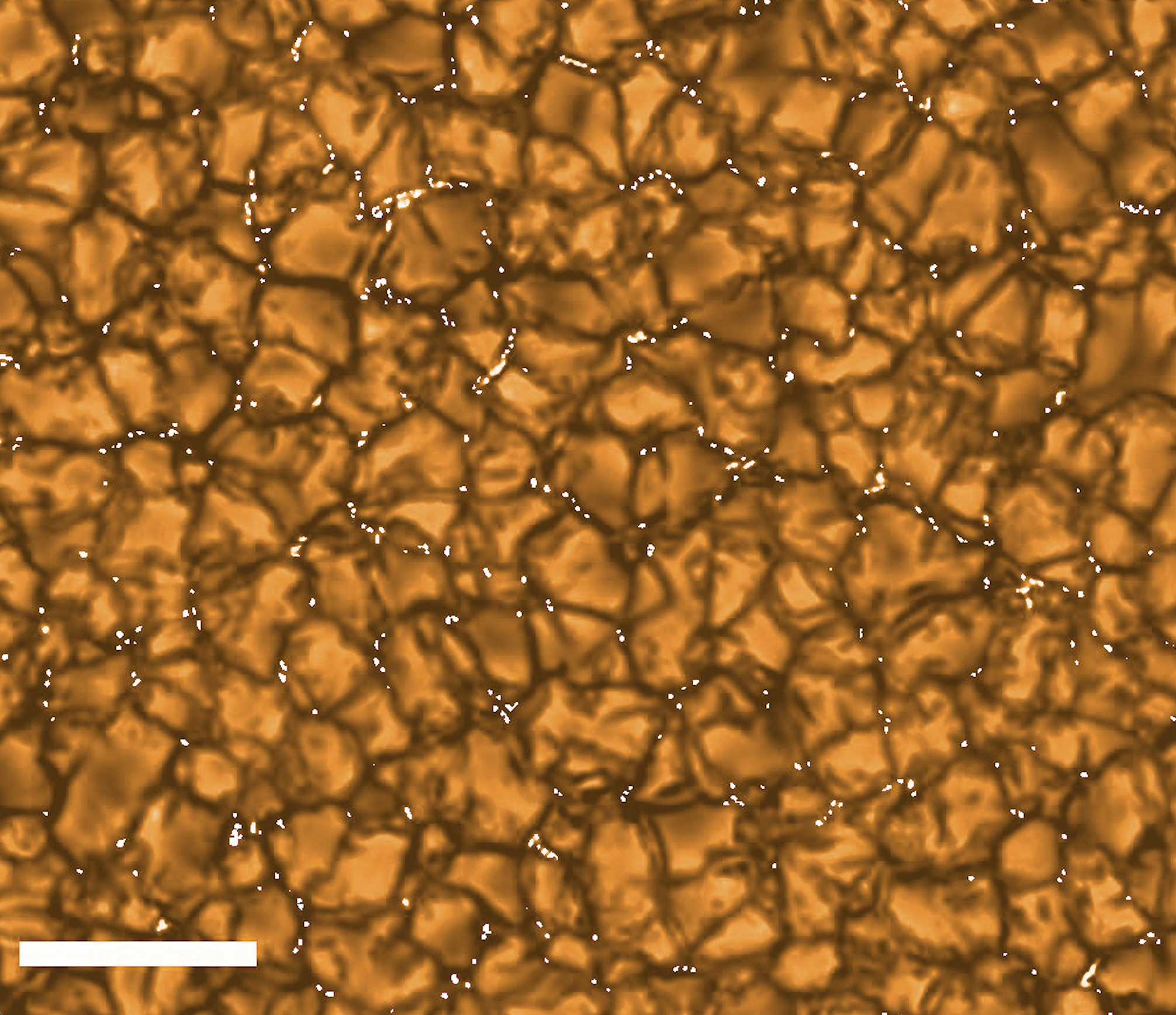}
\caption{Upper panels  display   the Sun's strongly variable facular coverage
  throughout its $24^{\mathrm{th}}$ magnetic cycle.
  The upper-left and upper-middle panels correspond to epochs near the Sun's  magnetic maximum, while the upper-right panel is at its magnetic minimum.
  Faculae are plotted in red, sunspots' umbra are plotted in black
  but are barely visible inside the active penumbra regions (plotted in blue).
  The lower panels illustrate the observational duality of faculae.
  The lower-left panel highlights their appearance as  brightened granules,
  whereas the lower-right panel reveals them as point-like,  inter-granular 
  G band bright points.
}
\label{f_2_fac_spo}
\end{figure}

To  illustrate the  profound differences  between facular  and sunspot
filling  factors throughout  a  solar activity  cycle,  we once  again
utilise reconstructions of the solar surface derived from the SATIRE-S
models, as detailed by \citet{yeo2014}.
The 2014 and 2015 snapshots in Fig.~\ref{f_2_fac_spo} capture the Sun
near solar maximum, whereas the  2020 snapshot corresponds to a period
of deep solar minimum.
From  left to  right, these  three upper  panels highlight  the highly
variable, expansive  surface coverage  of the  bright faculae  and the
magnetic network (shown in red). In stark contrast, sunspot umbrae are
plotted in black but are  barely visible within the active regions---a
visually  striking detail  that  perfectly  underscores their  minimal
spatial contribution---while the surrounding penumbrae are depicted in
blue.

To put  this striking  visual disparity  into a  quantitative context,
long-term measurements  confirm the overwhelming dominance  of faculae
over  sunspots.
An  analysis  spanning  the   entire  24$^{\mathrm{th}}$  solar  cycle
\citep{cretignier2024}  has shown  that while  the average  fractional
coverage    of   sunspots    ($f_{\rm   spot}$)    peaked   at    just
${\sim}0.1$$\%$  during maximum  magnetic activity,  the corresponding
plage    and    faculae    coverage    ($f_{\rm    plage}$)    reached
${\sim}3$$\%$---approximately $30$ times greater.
Other studies report varying  faculae-to-sunspot area ratios, with the
lowest  estimates being  ${\sim}10$  (\citealt{penza2024}) and  intermediate
values              falling               around              ${\sim}20$
(\citealt{shapiro2014,chatzistergos2022}).
Taken together, these findings  consistently indicate that the facular
disc  coverage  is typically  $10$  to  $30$  times  larger than  that  of
sunspots.

Notably, this  dramatic modulation in surface  coverage is intricately
linked to a striking transformation in the Sun's global magnetic field
configuration  over  the course  of  a  magnetic cycle.
During  solar   minimum,  the  magnetic  field   adopts  a  relatively
 `simple'     dipolar     structure     (\citealt{yoshida2023}),
characterised  by  a  well-defined   open  flux  that  extends
predominantly from  the polar  regions.
In  stark  contrast, at  solar  maximum,  this magnetic  configuration
undergoes a profound  shift: the polar contribution  weakens, and 
  low-latitude active regions  emerge, giving rise to  a more complex
and widely distributed  open field extending into interplanetary
space.

\subsection{Properties of faculae and magnetic bright points}
\label{s_faculae_dual_nature}
`Faculae' is  notoriously a loose  term in the literature.
While it strictly refers  to extended brightened granules, it
is frequently used as a generic catch-all for any  bright spot 
on the  photosphere.

To resolve this ambiguity, one must  examine the solar surface at high
angular  resolution. At these scales, the  quiet photosphere is
visibly  dominated by  solar  granulation:  convective cells  spanning
several  thousand   kilometres  with  lifetimes  of   up  to  ${\sim}20$
minutes.
In visible light, the borders of  a single granule appear dark because
the down-flowing plasma within  the inter-granular lanes is relatively
cooler  (by a  few hundred  degrees)  and therefore  presents a  lower
contrast compared to the hotter granular centre.
However,  observing  the   photosphere  through  specific  narrow-band
filters  reveals a  dramatically  different  picture. Embedded  within
these dark inter-granular lanes are small-scale intensity enhancements
known  as  photospheric  bright  points, magnetic  bright  points,  or
 G band       bright       points       (hereafter       GBPs;
\citealt{hao2020}).
GBPs  trace kilo-Gauss concentrations  of vertical
magnetic  flux  (\citealt{utz2013})  that  emit strongly  in  the  visible
continuum  (such as  the G band  and  TiO band)  and in  the wings  of
certain    spectral    lines    (e.g.    H$\alpha$    and    Ca\,{\sc
  ii}\,H$+$K).
Individual GBPs are tiny, typically  limited to the physical extent of
the  magnetic element,  spanning  $200$$-$$300$\,km or  less. Ensembles  of
these elements coalesce  to form a larger  structure commonly referred
to   as   the     photospheric   network   (\citealt{muller1985}),
 facular    points   (\citealt{mehltretter1974,muller1983}),    or
 filigree (\citealt{dunn1973}).

While GBPs are localised to the inter-granular lanes,  faculae 
(in the strictest sense) manifest on the granules themselves.
Although  both phenomena  share the  exact same  physical origin  (the
alteration  of  atmospheric  opacity  by magnetic  flux  tubes)  their
visibility, dynamics, and  brightening mechanisms differ fundamentally
(\citealt{berger2007}).
Faculae               are              larger               (averaging
${\sim}400$\,km in radial  width) and become visible  when the reduced
opacity within  a magnetic element  allows us  to see deeper  into the
hot, bright walls of an adjacent granule.
Interestingly,  the facular  component  has  a
characteristic  lifetime  of only  ${\sim}9$  minutes.  Because this  is
significantly shorter than the  ${\sim}20$-minute lifetime of a granule,
a remarkable property emerges:  multiple faculae can successively form
and dissipate on the very  same granule. Consequently, it is estimated
that  the  quiet  Sun  actually   hosts  more  faculae  than  granules
(\citealt{sanchez2010}).

Observationally,  the classical  distinction is  that single  GBPs are
predominantly  seen at  the solar  disc centre,  whereas ensembles  of
brightened  granules (faculae)  become highly  visible when  projected
near  the solar  limb. However,  there is  no strict  critical viewing
angle  at  which a  magnetic  brightening  cleanly switches  from  one
classification  to   the  other   (\citealt{berger2007}).
Indeed,  high-resolution comparisons  of G band  images with
Fe\,{\sc  i} $630.25$\,nm  Stokes\,$V$  magnetograms demonstrate  that
facular-like brightenings can  appear on granular edges  even near the
disc                                                            centre
($\mu$$=$$\cos$$\theta$$\ge$$0.8$),  likely  due  to  highly  inclined
magnetic  fields.   Conversely,  GBPs   can  still  be  identified  at
sufficiently limb-wards positions ($\mu$$\le$$0.6$).

Ultimately,  obscuration  and   dynamic  perspective  effects  dictate
whether  a  magnetic  element  appears  as a  GBP  or  a  facula.   As
highlighted  by  \citet{karoff2012},  these  are  inextricably  related
phenomena tied  to the  same magnetic  network. This  unified physical
relationship has  been robustly  confirmed across  direct observations
(\citealt{keller1992,berger2001}),  MHD simulations
(\citealt{schussler2003,carlsson2004,shelyag2004}), and semi-empirical
flux tube models (\citealt{steiner2001,sanchez2001}).

In this work, we use the  term faculae in its broader sense
  to  denote  both  the  photospheric faculae  clustered  near  active
  regions and those  widely dispersed across the quiet  Sun within the
  photospheric network.
To   visually   clarify   this    concept,   the   lower   panels   of
Fig.~\ref{f_2_fac_spo}   perfectly   illustrate  this   observational
duality.
Specifically, the lower-left panel  displays a high-resolution $430$\,nm
G band image (taken by the  Daniel K.\ Inouye Solar Telescope; credit:
NSO/NSF/AURA)  highlighting  the  appearance of  faculae  as  extended
 brightened granules when the Sun is viewed away from its disc centre.
Conversely, the  lower-right panel  presents a $430$\,nm  G band image
(taken  by  the  Swedish  Solar Telescope,  adapted  from  Fig.~1  of
\citealt{sanchez2010}),  demonstrating how  these magnetic  structures
appear as  point-like, inter-granular G band  Bright Points when
viewed predominantly at disc centre.
For  scale, the  white  bar  in both  lower  panels  corresponds to  a
distance of $5000$\,km across the Sun's surface.

\section{Numerical setup and parameters for magnetically
  enhanced MHD simulations}
\label{s_details_simul}

It is important to recall that faculae and spots always go hand
  in hand.  Therefore, any realistic simulation of small-scale
magnetic faculae must also incorporate the presence of large-scale
magnetic spots. However, we anticipate that modelling the 
  simultaneous presence of both features requires a dedicated setup,
distinct from that used to simulate faculae in isolation.
The  two complementary  simulations presented  here represent  a novel
endeavour;   there  are   no  established   frameworks  or   standard
methodologies for such an approach.

{\sc Faculae}: To evaluate the impact of an enhanced magnetic activity
and faculae presence on the CH and CN indices, we conduct 3D radiative
MHD simulations using the MURaM code (\citealt{vogler2005}).
The  MURaM  code  is  well-suited for  modelling  stellar  near-surface
convection  in the  presence of  magnetic fields,  as it  captures the
essential physics of solar facular regions. In this study, we employ a
MURaM  quiet-region   simulation  for  a  G2-type   dwarf  model  with
[M/H]$=$$-1$,  a  value  representative  of  the  metallicity  of  a
template globular cluster such as NGC104.
This metal-poor model was generated using the  small-scale dynamo 
(SSD) setup (\citealt{witzke2023}).

The faculae  simulation begins  with an  initial   seed--a uniform
vertical                magnetic               field                of
$200$\,G--which is then allowed to evolve dynamically. The choice of
this  absolute seed strength is somewhat arbitrary, as varying it
would  primarily influence  the  fraction of  the  surface covered  by
magnetic flux tubes (\citealt{beeck2015}).
Specifically,  a  stronger  initial  field would  increase  flux  tube
coverage,  thereby enhancing  the expected  impact  on the  CH and  CN
indices.               Notably,              the               assumed
$200$\,G field may also lead to the formation of pores--dark regions
resembling spots but lacking penumbrae--within the simulation domain.
The   simulation   runs   for    several   stellar   hours,   allowing
convection-driven magnetic  fields to organise into  regions of strong
vertical magnetism within intergranular  lanes. These regions form the
core  structures of  faculae, often  idealised  as flux  tubes in  the
literature. Upon completion, we compute the CH and CN index variations
by comparing:
(i)                                                                the
$200$\,G   small-scale   dynamo   (SSD)     seed   simulation,
representing a quiet stellar region,  and (ii) the same SSD simulation
after the evolution of faculae.
This  comparative analysis  is carried  out using  the MPS-ATLAS  code
(\citealt{witzke2021}),   operating  in   its     1.5-D   mode--a
ray-by-ray  approach  (\citealt{witzke2024})   that  enables  detailed
spectral synthesis across the simulation domain.

To account  for higher levels  of surface magnetisation, we  scale the
estimated CH and CN index variations by a factor of $B/200$, where $B$
denotes the mean unsigned surface magnetic field.
Field strengths are allowed to  reach up to $3000$\,G--an extreme case
in which  the stellar surface  is   fully covered by  flux tubes.
While clearly unrealistic  for our Sun, the  maximum possible coverage
by small-scale flux tubes remains uncertain.
Nevertheless,  a portion  of  the simulated  magnetic  flux will  
  inevitably organise  into  large flux  tubes, giving  rise to
spots.  This unavoidable  coupling  underscores the  need to  simulate
faculae and spots  simultaneously.

{\sc  Spots}: In  light  of  the above  considerations,  we perform  a
dedicated,   parallel,  and   complementary   simulation  focused   on
starspots.  This scenario  models a  linear increase  in surface  spot
coverage with increasing input magnetic field strength $B$.
Specifically,    surface    coverage    ranges   from    $0$$\%$    at
$B$$=$$0$\,G              to              $50$$\%$              at
$B$$=$$3000$\,G. In  this upper-limit  case, the stellar  surface is
evenly divided: $50$$\%$ covered by spots and the remaining
$50$$\%$ by small-scale flux tubes (faculae).
It is  important to note that  MHD simulations of spots  in stars with
non-solar metallicity  are not  yet available.   However, it  has been
shown (\citealt{smitha2025}) that 1D models can reliably estimate spot
spectra in G2-type dwarf stars.
Accordingly,  we  determine  spot  contrast  using  a  1D  model  grid
(\citealt{kostogryz2022,kostogryz2023}), where  spots are approximated
by  stellar  atmospheres that  are  $2000$\,K  cooler than  the  quiet
photosphere.

\section{The rotation-magnetism link in young and
  intermediate-age clusters}
\label{s_young_clusters}

A   crucial  observational   constraint  for   any  framework
  addressing the MP  phenomenon is provided by the study  of young and
  intermediate-age stellar  clusters with  masses   comparable to
  those of old GCs \citep[see e.g.][]{bastian2018}.
Over the past decade, extensive photometric and spectroscopic
  surveys   have    established   a   distinct   age    threshold   of
  $\sim$$2$\,Gyr,  below  which  the   light-element  spreads
  \citep[characteristic  of old  GCs;][]{carretta2009}  are no  longer
  detected.
  This  age  limit is  robustly  supported  by  a sequence  of  recent
  systematic  investigations.  Indeed,  photometric  evidence  of  the
  absence of MPs  in clusters younger than this  threshold was clearly
  demonstrated   by  \cite{martocchia2018},   a  result   subsequently
  extended to  the spectroscopic  domain through the  documentation of
  homogeneous     nitrogen     distributions      (the     lack     of
  $\mathrm{N}$-spreads) by \cite{martocchia2021}.
  Further    comprehensive     analysis    by    the     same    group
  \citep[e.g.][]{saracino2020} have solidified this temporal boundary,
  while recent  deep photometry  by \cite{cadelano2022}  confirms that
  this   structural  homogeneity   extends  down   to  the   unevolved
  main-sequence stars.

  This                                                       empirical
  $\sim$$2$\,Gyr  boundary directly  relates  to how  a cluster's  age
  determines the mass of stars currently at the turnoff, which in turn
  governs the core-envelope architecture of its evolving members.
Specifically, the currently surviving stars near the 
turnoff of a $\le$$2$\,Gyr young cluster are higher-mass stars 
($\gtrsim$$1.5$M$_{\odot}$) that possess radiative envelopes. Conversely, 
past this progressive structural transition zone, the currently surviving 
stars of a $\ge$$2$\,Gyr older cluster are lower-mass stars 
($\lesssim$$1.2$M$_{\odot}$) characterised by convective envelopes.
This foundational boundary in stellar astrophysics is known as the Kraft break 
\citep{kraft1967}.

Spanning  the continuous  structural  transition zone  around
  $\sim$$1.5$M$_{\odot}$                                      (T$_{\rm
    eff}$$\sim$$6250$\,K  in  main-sequence  stars), it  represents  a
  critical structural and dynamical boundary:
(i) main-sequence stars with extremely thin or non-existent surface 
convective zones (i.e. $\gtrsim$$1.5$M$_{\odot}$) cannot drive a 
solar-like dynamo, thereby retaining their rapid primordial rotation 
velocities throughout their main-sequence lifetimes as  fast rotators 
\citep[see also][]{vansaders2013}, whereas
(ii)   stars  possessing   substantial   convective  envelopes   (i.e.
$\lesssim$$1.2$M$_{\odot}$)  drive an  internal  magnetic dynamo  that
couples  with stellar  winds. This  global configuration  continuously
strips angular momentum from the star, effectively exerting a magnetic
braking  torque \citep{mestel1968,  skumanich1972} that  significantly
spins down the stars as they age, ensuring they remain slow rotators.

While  deep  internal  mixing   mechanisms  (such as  the  first
  dredge-up) and associated evolutionary  caveats must be meticulously
  weighed    when     evaluating    these     population    boundaries
  \citep{salaris2020}, this observational  cutoff clearly implies that
  the  mechanism  driving  spectroscopic   anomalies    is  not  a
    universal property of all stellar aggregates.
Instead, the MP phenomenon is strictly governed by the intrinsic
structural or dynamical properties of the individual stars themselves.

The striking empirical coincidence between the turnoff mass dictated by 
this $\sim$$2$\,Gyr age threshold and the structural boundary 
separating fast from slow-rotating main-sequence stars strongly hints 
that the spectroscopic anomalies are physically linked to stellar 
rotation and surface magnetic fields.
From a \textsc{MAGISTER FAS} standpoint, the disappearance of
  abundance               anomalies               at               the
  $\sim$$1.5$\,M$_{\odot}$ boundary emerges naturally,  as it traces the
  underlying transition  in the structural engine  that powers surface
  magnetism.
  This  physical mapping  is strongly  supported by  long-term stellar
  activity  surveys. Indeed,  by  contextualising modern  observations
  from the  TIGRE telescope  within the historic  legacy of  the Mount
  Wilson chromospheric activity monitoring project, \citet[][see their
  Fig.~15]{schmitt2026} confirm that  robust, cyclic magnetic activity
  operates across the entire main-sequence mass range characterised by
  convective    envelopes,     extending    right    up     to    this
  $\sim$1.5\,\text{M}$_{\odot}$ threshold.
Ultimately, the seamless capacity of a framework anchored in
$\sim$$5$\,Gyr solar physics to bridge the critical observational
 `switch' at the younger $\sim$$2$\,Gyr threshold with the
anomalous properties of ancient ($>$$10$\,Gyr) GCs further
bolsters its overall viability.

\section{Stellar activity cycles versus the
  temporal stability of multiple populations}
\label{s_magnetic_cycles}

Having anchored  the physical basis of  our proposed scenario
  in the solar magnetic cycle, it  is necessary to address whether the
  magnetic activity typical of solar-like stars could induce transient
  spectroscopic  migrations,  effectively   causing  member  stars  in
  globular clusters  to transition  between 1P and  2P classifications
  over decadal timescales.
  Two  complementary  lines  of   empirical  evidence  from  long-term
  monitoring programmes  of Galactic  field stars, when  taken together,
  strongly suggest otherwise.

\begin{figure}
\centering
\includegraphics[width=0.75\hsize]{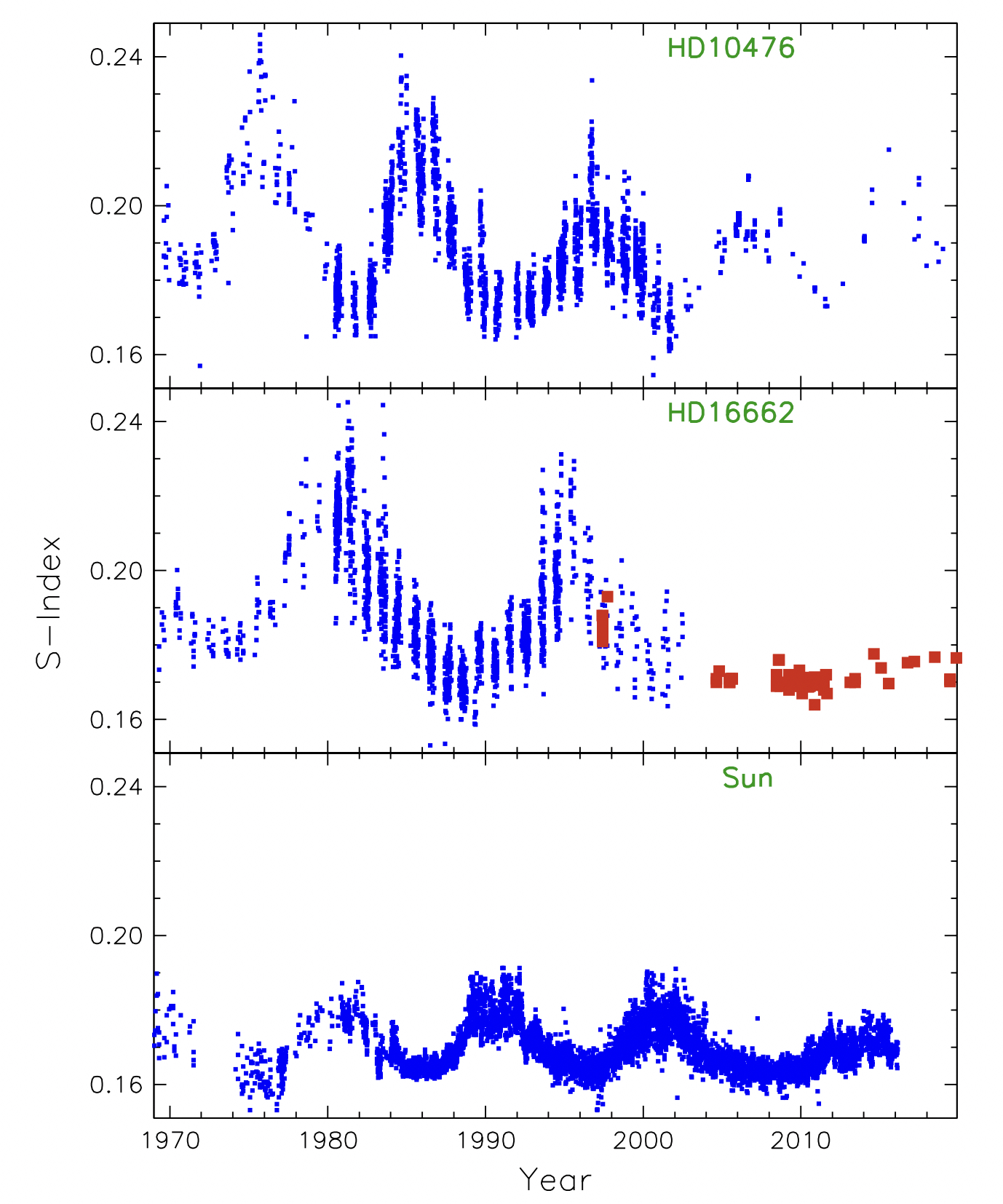}
\caption{
  Comparative multi-decadal chromospheric activity tracking.
  The upper panel  shows $\sim$$50$\,years of Ca{\sc ii}\,H$+$K S-index
  monitoring \citep{baum2022}  for the solar analogue HD10476
  [T$_{\rm eff}$$=$$5190$\,K, M$=$$0.78$\,M$_{\odot}$,  [Fe/H]$=$$-0.03$,
  age $\sim$$8.1$\,Gyr], which exhibits a well-defined  $\sim$$10.3$\,year
  magnetic cycle.
  The middle panel presents similar S-index   monitoring
  \citep{baum2022}           for           HD166620           [T$_{\rm
    eff}$$=$$4970$\,K,                             $\log\,g$$=$$4.51$,
  M$=$$0.76$\,M$_{\odot}$,           [Fe/H]$=$$-0.16$,           age
  $\sim$$12.4$\,Gyr],      displaying      two     fully      resolved
  $\sim$$17$\,year cycles.
  Data points marked with red symbols \citep{luhn2022} firmly confirm that this 
  star has transitioned into a prolonged, Sun-like Maunder minimum state. The 
  lower panel displays the long-term solar S-index monitoring
  data \citep{egeland2017}  for comparison.
  The Y-axis scale is strictly identical across all three panels, 
  illustrating both the relatively quiescent nature of the Sun and the long-term 
  stability of individual stellar activity ranges.
}
\label{f_3stars}
\end{figure}

First,  chromospheric surveys—pioneered  by the  Mount Wilson
  stellar   activity  programme   \citep{wilson1968}  and   extended  by
  subsequent      studies      \citep{baliunas1990,      baliunas1995,
    schmitt2026}—have established that magnetic  activity cycles are a
   ubiquitous feature among  low-mass stars possessing convective
  envelopes.
  To    visually    contextualise     this    fundamental    property,
  Fig.~\ref{f_3stars} presents a comparative overview of multi-decadal
  Ca\,{\sc  ii}\,H$+$K  $S$-index\footnote{The Mt.\  Wilson  $S$-index
    \citep{vaughan1978, wilson1978}  quantifies chromospheric activity
    as the ratio of the summed  fluxes in the Ca\,{\sc ii}\,H$+$K line
    cores to two nearby  pseudo-continuum reference bands.} monitoring
  data  for two  well-studied low-mass  field analogues  alongside the
  Sun.
As illustrated in the upper panel, HD10476 represents the classic cyclic 
regime, displaying a highly regular, stable $\sim$$10.3$\,year magnetic 
cycle over a $\sim$$50$\,year observation baseline.
In particular,  HD10476 exhibits a highly  pronounced stellar activity
cycle      with     a      peak-to-trough     $S$-index      amplitude
($\Delta$$S$$\sim$$0.070$)          that           is          roughly
$2.5$   times  larger   than   the  typical   solar  cycle   variation
($\Delta$$S$$\sim$$0.028$).
On the  other hand, the  middle panel demonstrates that  these stellar
dynamos can operate on even longer temporal scales, with the old dwarf
HD166620         displaying         two         fully         resolved
$\sim$$17$\,year cycles \citep{baum2022}.
Crucially, when viewed  on a strictly identical  vertical scale, these
two  records   confirm  that  solar  analogues   consistently  sustain
well-defined, multi-decadal  activity cycles—directly  reinforcing the
universal ubiquity of these dynamos—while simultaneously demonstrating
that the Sun  operates on a relatively quiescent  baseline compared to
its field peers \citep[see also][]{schmitt2026, reinhold2020x}.

To  firmly establish  that  such cyclic  behaviour  represents a
 universal  property rather  than an  isolated phenomenon,  a recent
comprehensive  analysis by  \cite{isaacson2024} examined  a sample  of
$285$ stars  with sufficient temporal  sampling to detect  cycles with
periods                          ranging                          from
$2$$\div$$25$\,years, confirming  periodic   magnetic modulations
in $138$ systems.
  Crucially,     in     the     T$_{\rm    eff}$     range     between
  $4700$$-$$5900$\,K [the regime  directly relevant to  unevolved dwarf
  stars  in  GCs  spanning  from the  main-sequence  turnoff  down  to
  $\sim$$3$\,magnitudes  below] nearly  every   single  monitored  star
  displays a regular magnetic cycle.
  This   analysis   also   confirms  that   magnetic   cycle   periods
  systematically increase as effective temperature decreases: yielding
  $4.4$$\pm$$0.5$\,years       within        the       solar       bin
  ($5600$$\div$$5900$\,K),
  $6.0$$\pm$$0.7$\,years
  ($5300$$\div$$5600$\,K),
  $7.2$$\pm$$1.1$\,years
  ($5000$$\div$$5300$\,K),                                         and
  $7.8$$\pm$$2.0$\,years        at        the       coolest        end
  ($4700$$\div$$5000$\,K).

  There is  no physical  basis to  expect that  low-mass main-sequence
  stars  in globular  clusters  would behave  differently. Similar  to
  their field counterparts, they  must exhibit similar magnetic cycles
  and associated secular trends.

  Second, secular trends operating on  top of periodic magnetic cycles
  do not translate into stellar `instability'.  Supporting evidence of
  this  is provided  by  the Sun  itself.  Indeed, historical  sunspot
  records over
  the                                                             past
  $\sim$$400$~years   demonstrate   that    it   naturally   undergoes
  multi-decadal variations characterised by  extended grand maxima and
  minima \citep{penza2024}.
  In a similar vein,  Total Solar Irradiance  (TSI) reconstructions
  over the
  past five  centuries confirm  that the  solar dynamo  can transition
  into  prolonged phases  of relative  inactivity, exemplified  by the
  $\sim$$70$-year  Maunder  minimum  \citep{penza2022}.
  On  much  longer  timescales, reconstructing  sunspot  numbers  from
  $^{14}$C data allowed  \citet{usoskin2007} to isolate a  total of $27$
  solar     `grand     minima'     episodes    over     the     last
  $11300$~years,   revealing    that   the   Sun    spends   roughly
  $17$$\%$ of  its lifetime in  such a dormant state.
  Crucially,    the   Sun    navigates    these   dramatic    magnetic
  transitions—shifting  between  normal  cyclic regimes  and  extended
  minima—in  a  highly  regular fashion  \citep{schmitt2026}.
  These  transitions  exert  a negligible  impact  on  global
  stellar  parameters;  variations  in   total  energy  output  remain
  strictly confined to the sub-percent level, ensuring the fundamental
  stellar identity remains unchanged.

While  capturing a  similar phase  of relative  inactivity in
  other stars has long represented an elusive, decades-long challenge,
  the persistent tracking of HD166620 (cf.\ red symbols in the middle
  panel  of Fig.~\ref{f_3stars})  has  ultimately  provided the  first
  definitive confirmation of an extrasolar analogue transitioning into
  a true Maunder minimum state \citep{baum2022, luhn2022}.
  Remarkably,   despite   the    profound   magnetic   reconfiguration
  experienced  during this  transition, HD166620  has maintained  its
  phase  of inactivity  without   altering  its overall  optical
  luminosity \citep{luhn2022}. This robustly  demonstrates that even a
  complete  suppression of  cyclic  activity  represents a  temporary,
  topological  reconfiguration   of  surface  fields  rather   than  a
  structural  internal alteration  capable of  destabilising a  star's
  identity.

The  observed  luminosity   and  thermodynamic  stability  of
  HD166620—consistent with  our own Sun—reinforce the  core premise of
  the \textsc{MAGISTER FAS} framework:  cyclic magnetic modulations should 
  not drive cross-population transitions between  1P and 2P regimes in
  GCs. Instead, these  sub-populations are distinguished fundamentally
  by their different long-term  average magnetic baselines and surface
  field topologies that remain invariant over evolutionary timescales.
  Consequently,  while  individual  stars  undergo  periodic  activity
  cycles, these  variations act  merely as minor  perturbations around
  the  star's stable,  intrinsic  mean. This  effectively anchors  the
  spectroscopic profile of each star to its baseline identity, thereby
  precluding   any  1P/2P   population  reclassification   over  human
  observation timescales.

  \section{Surface magnetism and atomic abundance
    indicators in dwarfs and giants}
\label{s_atomic_variations}

The physical principles  underlying the \textsc{MAGISTER FAS}
  framework  imply that  the atomic  transitions (historically  used to
  define  cluster populations)  must  also be  subject to  magnetically
  induced   line   profile   variations,  affecting   both   unevolved
  main-sequence (MS) and evolved red giant branch (RGB) stars.
  To ground  these theoretical  expectations in empirical  reality, we
  first utilise the high-resolution diagnostic potential of the Sun to
  examine the implications for cluster MS stars.

  \subsection{Solar benchmark and main-sequence dwarfs}
\label{ss_ms_magnetism}

A recent  ultra-high-resolution study  \citep{pietrow2026} of
  the  solar  oxygen  infrared  triplet   has  confirmed  a  weak  but
  significant  variation  across  the activity  baseline.  While  this
  signature is small  enough that the inferred  solar oxygen abundance
  remains stable  throughout the  solar cycle,  the authors  note that
  this effect may become significant for more active stars.
  This  expectation   is  robustly  supported  by   decades  of  solar
  monitoring  \citep{livingston2007}, which  reveal that  the physical
  state of the stellar  atmosphere is continuously reconfigured across
  different  optical depths.
  Specifically, in the high photosphere, strong spectral features—such
  as the core profiles of \ion{Fe}{i}, the \ion{Na}{i}~D1/D2 resonance
  doublets,  and  the  \ion{Mg}{i}~b lines—display  complex,  volatile
  modulations that mirror the $22$-year  Hale magnetic cycle rather than
  simple   $11$-year  variations   \citep[cf.][Fig.~17]{livingston2007}.
  Concurrently,   long-term   tracking   of   weak,   deep-photosphere
  transitions,  such as  the \ion{Fe}{i}~537.9,  \ion{C}{i}~538.0, and
  \ion{Ti}{ii}~538.1\,nm   lines   \citep{penza2006},   has   unveiled
  distinct   periodicities  corresponding   to   both  the   classical
  $\sim$$11$-year cycle and its shorter $\sim$$2.8$-year harmonic.
  It is  therefore highly plausible  that the solar  dynamo, operating
  through  both  multi-decadal   and  short-term  magnetic  harmonics,
  continuously influences line-formation  parameters.
  Crucially,  these  weak   photospheric  transitions,  while  largely
  unaffected  by sunspot  flux-blocking,  exhibit a  subtle but  clear
  sensitivity  to  bright facular  configurations  \citep{penza2006}—a
  dependency  that  remains  entirely  unexplored in  the  context  of
  globular cluster stellar architectures.

Interestingly, this complex atmospheric behaviour---\allowbreak manifesting
as a systemic, magnetically driven thermodynamic perturbation---is
naturally explained by the localised facular `hot wall' scenario
\citep{spruit1976, solanki2013}.
  For instance,  direct
  analysis of effective spectral  line temperature indicators spanning
  more than a solar cycle  unambiguously demonstrates that the average
  solar temperature  varies systematically and in  phase with magnetic
  activity  by  $+1.5$$\pm$$0.2$\,K  \citep{gray1997}.
  This variation aligns with earlier investigations demonstrating that
  mid-photosphere lines unequivocally  register a systemic temperature
  rise                                                              of
  $+$$8$\,K between activity minimum and maximum \citep{mitchell1991}.
  Rather  than reflecting  a  uniform heating  of  the global  stellar
  surface,  these   observed  temperature  variations   represent  the
  disc-integrated manifestation of millions of evacuated magnetic flux
  tubes  distributed across  the  active Sun.
  At  peak activity,  the  viewing geometry  of  these bright  faculae
  exposes  their  deeper,  hotter  granular  walls  to  the  observer,
  systematically shifting the thermodynamic  state of the line-forming
  layers.
  Thus,  across independent  instrumentation, diagnostic  methods, and
  epochs,  the empirical  consensus remains  robust: elevated  surface
  magnetic coverage consistently drives  a measurable temperature rise
  within    the    photosphere,    inevitably    inducing    systemic,
  magnetically-driven thermodynamic perturbations.

When   a  spatially   unresolved,  disc-integrated   spectrum
  containing  these  multi-component, thermodynamically  altered  line
  profiles  is modelled  using  a standard  1D  LTE radiative  transfer
  code—which  assumes  a  completely static,  flat,  and  unmagnetised
  atmosphere—the  code  cannot  physically  interpret  the  underlying
  profile  variations.
  Instead,   conventional   fitting  procedures,   whether   utilising
  equivalent widths  or spectral  synthesis, may compensate  for these
  subtle  profile modifications  by  shifting  the inferred  elemental
  abundances, potentially mapping purely thermodynamic variations onto
  apparent, spurious chemical spreads.

Whether this mechanism can consistently reconcile the diverse
  abundance indicators  observed across different  cluster populations
  remains  an open  question. Fully  exploring this  hypothesis across
  various  evolutionary stages  represents a  demanding frontier  that
  will  ultimately require  dedicated  3D  non-LTE magnetic  radiative
  transfer modelling.

\subsection{Scaling  the framework  to  the red  giant
  branch}
\label{ss_rgb_magnetism}

In parallel to the potential implications of magnetically induced line
alterations  in  MS  cluster  stars, an  important  open  question  is
whether,  and to  what extent,  such variations  might affect  evolved
stars along the RGB.
Indeed, spectro-polarimetric surveys of low-mass evolved stars 
\citep[e.g.][]{auriere2015} confirm that surface magnetic fields remain 
ubiquitous on the RGB.
However,   while  global,   disc-averaged  magnetic   field  strengths
naturally           dilute           to           low           values
($\sim$$0.5$--$5$\,G)   due  to   substantial  radius   expansion  and
rotational  braking,  the  resulting  effect on  line  profiles  could
theoretically be enhanced—rather  than mitigated—within these extended
atmospheres.
This  potential  enhancement  is  expected   to  be  driven  by  three
cooperative physical mechanisms.

First,  RGB stars are significantly 
  cooler                                                      (T$_{\rm
    eff}$$\sim$$4000$--$4800$\,K)    than    main-sequence    turn-off
  stars. At these lower temperatures,  the chemical equilibrium of key
  molecular        tracers       ($\mathrm{CH}$,        $\mathrm{CN}$,
  $\mathrm{NH}$,                                                   and
  $\mathrm{OH}$) is highly non-linear; consequently, a minor localised
  thermal perturbation within a weak  magnetic element could trigger a
  significant fractional change in molecular dissociation, potentially
  amplifying     the    observed     apparent    abundance     spreads.
  Second, the  low atmospheric densities characteristic  of red giants
  ($\log
  g$$\lesssim$$2.5$) are  well known  to induce departures  from Local
  Thermodynamic Equilibrium. Because conventional 1D hydrostatic
  radiative transfer  analyses are structurally blind  to both non-LTE
  population effects and the  multi-component 3D convective structures
  inherent   to  magnetised   giant   atmospheres,  standard   fitting
  procedures  may   map  these   subtle  profile   modifications  into
  artificial chemical spreads.

  Third,  because  a  star's tidal  interaction  cross-section  scales
  strongly  with  its  radius,  inflated red  giants  represent  prime
  targets  for  stochastic  angular momentum  injection  during  close
  stellar encounters in dense cluster cores.
  This sporadic dynamical
  interaction could  provide a mechanism for  rotational rejuvenation,
  potentially helping to sustain active surface dynamos.

  Overall, the physical processes underlying the \textsc{MAGISTER FAS}
  framework might remain  highly relevant across the  RGB phase, where
  they could theoretically operate with heightened impact.

\section{The 1P/2P dichotomy among M dwarfs in globular clusters}
\label{s_mtype}

The  photometric  splitting of  the  MS  in  globular
clusters  is  a  remarkably   persistent  feature,  spanning  a  broad
temperature       range       from       late       F-type       stars
(${\sim}6500$\,K) with radiative cores down to fully convective M-type
stars (${\sim}2500$\,K).
Traditionally,  the  persistent   splitting  observed  among  M dwarfs
\citep{marino2024} has  been interpreted  as evidence of   distinct 1P
and 2P stars  formed in sequence.
 This  conviction  rests on  the  premise  that because  fully
  convective                        M-type                       stars
  ($\lesssim0.35\,\mathrm{M}_{\odot}$)  cannot   experience  radiative
  levitation,  gravitational  settling, or  nucleosynthetic  dredge-up
  episodes,  their observed  surface abundances  must strictly  mirror
  their primordial compositions.  Consequently,  any observed split in
  the fully  convective regime is  canonically viewed as  an immutable
  chemical signature inherited at birth.

Here,  we explore  how the  photometric splitting  observed in  the
M-dwarf   regime  can   be  qualitatively   accommodated  within   the
\textsc{MAGISTER FAS} framework.
Specifically, we illustrate how  small-scale magnetic fields (faculae)
not only  can produce the  observed MS splitting in  F/G/K-type stars,
but can also sustain this signature throughout the M-type regime.
This  application of  our framework  is supported  by mounting
empirical  evidence  that   M-dwarfs  possess  inhomogeneous,  spotted
surfaces indicative  of facular  activity.  Indeed, M-type  dwarfs are
almost universally recognised  to display (spot-based) rotation-period
vs. magnetic-activity  relationships   akin  to  their  earlier-type
counterparts \citep{reiners2014, wright2016, lu2023}.

\begin{figure}
\centering
\includegraphics[width=0.99\hsize]{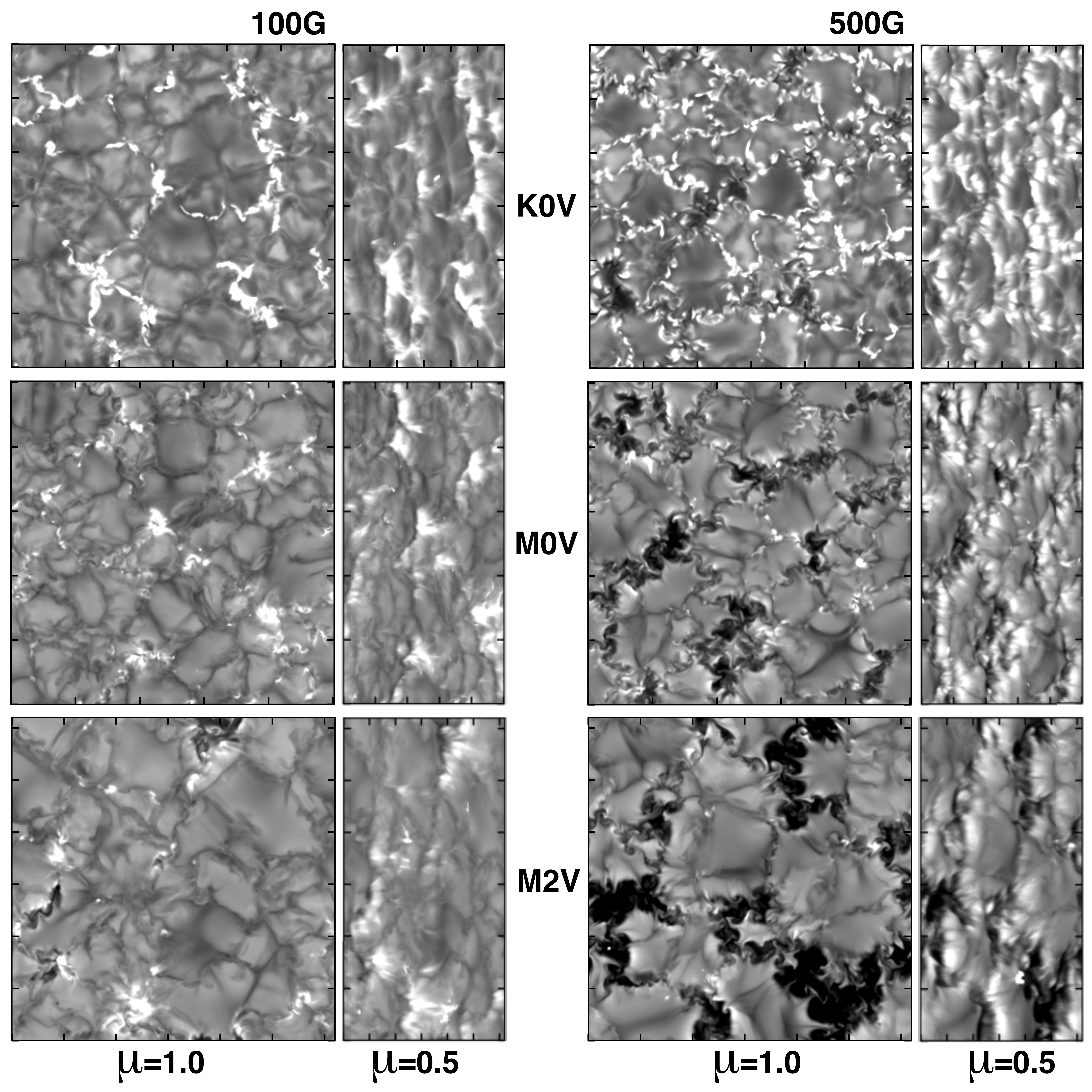}
\caption{
  Reconstructed 3D radiative MHD simulation results (adapted from 
\citealt{norris2023}), illustrating the structural changes induced by 
faculae across the K--M spectral sequence and highlighting the striking 
inversion from bright to dark features in cool M-type models.
  The two left  columns display emergent intensities  at $388$\,nm for
  an       average       magnetic        field       strength       of
  $\langle$$B$$\rangle$$=$$100$\,G, viewed  at: (i)   disc-centre 
  ($\mu$$=$$1.0$)     and    (ii)     the       stellar    limb 
  ($\mu$$=$$0.5$).  The two  right columns  display the  corresponding
  results                for                 a                stronger
  $\langle$$B$$\rangle$$=$$500$\,G magnetic field.
}
\label{f_app_2}
\end{figure}

First, to illustrate how faculae alter the thermodynamic structure of 
the photosphere and potentially trigger a MS duality in F/G/K-type 
stars, we rely on the 3D radiative MHD simulation results of 
\citet{norris2023}, reconstructed in Fig.~\ref{f_app_2}.
A comparison of the K0V  simulations at $100$\,G and $500$\,G provides
a visual  demonstration of  how a  relatively stronger  magnetic field
produces  a brighter  (hotter) stellar  surface.  This  effect remains
persuasive    across   all    viewing    angles,   from    disc-centre 
($\mu$$=$$1.0$) to the stellar limb ($\mu$$=$$0.5$).

%
%

%
Second, to illustrate how faculae can potentially trigger a MS duality 
within M-type stars, we again adapt the 3D radiative MHD simulation 
results of \citet{norris2023}, drawing the attention of the stellar 
community to a striking facular 
`inversion' property.
This effect is best appreciated when examining 
the $500$\,G simulations across the spectral range from K0V to M2V: 
the same $500$\,G field that generates bright faculae in the K0V model 
instead produces unexpectedly dark faculae in the cooler M0V and M2V 
models. Similarly, the M2V $500$\,G simulation induces significantly 
darker faculae than the corresponding $100$\,G M0V run.
This trend, 
reconstructed from the \citet{norris2023} simulation data and supported 
by independent literature studies \citep{steiner2014, beeck2015}, confirms 
that faculae undergo a fundamental bright-to-dark transition as 
effective temperature decreases.
Physically,  this   phenomenon  occurs   because  flux   tubes  become
increasingly  shallow  as  one  moves from  G-type  to  M-type  dwarfs
\citep{beeck2015,shapiro2026}.    Consequently,  shallow   flux  tubes
experience  less  radiative heating  from  the  `hot walls'  of  the
surrounding  granules \citep{solanki2013},  resulting in  magnetically
concentrated  regions  that  are  cooler  and  darker  than  the  mean
photosphere.
Stronger surface magnetism  in the M-type regime is  thus simulated to
produce a  differentially darker surface, potentially  contributing to
the observed MS splitting.

Interestingly, this  striking facular  transition provides  an elegant
qualitative  explanation  for   the  puzzling  1P/2P  `colour-switch'
observed    in   near-infrared    colour-magnitude   diagrams
\citep{milone2012x, milone2023,  cadelano2023, marino2024, milone2026}
at the transition to the fully convective M-dwarf regime.
In  particular,  within  the   \textsc{MAGISTER  FAS}  framework,  the
following characteristics would emerge:

\begin{itemize}
\item Above the transition: Magnetically `active' (2P-like)
  stars  are  expected  to  be   relatively   bluer   than  their
  `less-active' counterparts because stronger magnetic fields induce
  brighter and hotter facular regions.
\item Below  the transition:  The same  relatively `active'
  stars become   redder   because those  same magnetic  fields now
  induce darker inter-granular depressions, increasing surface opacity
  and shifting the integrated colour towards the red.
    \end{itemize}

    Third,  a  noteworthy  property  of M-type  dwarfs  identified  by
    \citet{milone2019}  is that  the  1P/2P  population ratios  remain
    essentially    identical   across    the    entire   mass    range
    (${\sim}$$0.15$--$0.80\,\mathrm{M}_{\odot}$).
    This finding stands in sharp contrast to canonical self-enrichment
    scenarios  \citep{ventura2001,  dantona2016},  which  require  the
    ad-hoc assumption  that GCs preferentially lost  the vast majority
    of their  primordial 1P stars  to the Galactic field  during early
    dynamical   evolution   \citep[i.e.  the   mass-budget   problem;
    e.g.][]{decressin2010, lacchin2024}.
    Within  the  \textsc{MAGISTER  FAS}  framework,  the  strict  mass
    invariance of  this ratio naturally arises  because the underlying
    physical  mechanism  establishing  the bimodal  magnetic  activity
    distribution  operates  uniformly   across  the  low-mass  stellar
    spectrum.  It bypasses  any requirement  for spatial  stripping or
    fine-tuned mass loss.

    In  conclusion, \textsc{MAGISTER  FAS} offers  a unified  physical
    pathway  that, in  principle,  simultaneously  accounts for  three
    distinct observational trends across the F/G/K and M-type regimes:
    (i) the continuous MS photometric  splitting, (ii) the characteristic
    P1/P2 near-infrared  colour-switch unique to  the M-dwarf
    domain, and (iii) the observed 1P/2P population ratios.
    
\section{Sensitivity of solar molecular indices
      to the magnetic cycle}
\label{s_7indici}

A primary advantage of the \textsc{MAGISTER FAS} framework lies in the
recognition   and   consensus   \citep[e.g.][]{sbordone2011,
    lee2019,  milone2020}  that  the photometric  sequence  splitting
observed  in   globular  clusters—detected   primarily  in   mid-  and
near-ultraviolet filters and only  marginally in the optical regime—is
inextricably linked to  the response of specific  molecular bands (OH,
NH, CN, and CH).
In particular,  these molecular  bands fall  directly within
specific  Hubble Space  Telescope  filter bandpasses  (e.g. 
$F275W$,   $F336W$,   $F343N$,   $F438W$)  customarily   employed   to
photometrically disentangle 1P and 2P populations.
The  consensus on  the primary  role  of molecular  band strengths  in
driving  the MP  photometric  signature has  motivated  the design  of
ground-based narrow-  and intermediate-band filter  systems explicitly
tailored to  trace the characteristic imprints  of the NH, CN,  and CH
features \citep[e.g.][]{lee2019}.

From a \textsc{MAGISTER FAS} perspective, there is a compelling solar 
physics parallel to the observed mid- and near-ultraviolet spectral 
dependence of MPs:
while the Total Solar Irradiance (TSI) varies by 
only $\sim$$0.1$$\%$ over a magnetic cycle, approximately $60$$\%$ of this 
variation is concentrated at wavelengths shorter than $400$\,nm 
\citep{mitchell1991, krivova2006, solanki2013}.\footnote{Figure\,3 in 
\citet{shapiro2016} vividly illustrates this ultraviolet dominance of 
solar irradiance variability, demonstrating that below $400$\,nm, 
these variations are almost exclusively faculae-driven.}
This alignment points to stellar magnetism as the underlying driver of 
the photometric signatures of MPs.
This interpretation is further reinforced  by the fact that these same
molecular indices in the Sun  exhibit a clear, synchronised dependence
on  the  magnetic  cycle,  ultimately  indicating  that  the  observed
sequence splitting in GCs represents a natural observational signature
of differential surface magnetism.

To evaluate this susceptibility, we analysed seven molecular
indices derived  from the SATIRE-S continuous  reconstruction of daily
solar  irradiance  \citep{yeo2014}.  Our  focus is  primarily  on  the
$23^{\mathrm{rd}}$ solar cycle (1996--2008), which provides a complete
and well-sampled view of molecular variability from minimum to maximum
activity.
All  molecular  indices  were  converted  into  magnitudes  using  the
standard logarithmic flux ratio:

\begin{equation}
\mathrm{Index} = -2.5 \log_{10} \left( \frac{F_{\mathrm{feature}}}{F_{\mathrm{continuum}}} \right),
\end{equation}

where  $F_{\mathrm{feature}}$  represents  the flux  in  the  specific
molecular window (e.g. CH4300, violet  CN4142, CN3883, NH ammonia, or
the  OH hydroxyl  bands) and  $F_{\mathrm{continuum}}$ represents  the
flux in the adjacent reference regions as defined in the text.

\begin{figure}
  \centering
\includegraphics[width=0.82\hsize]{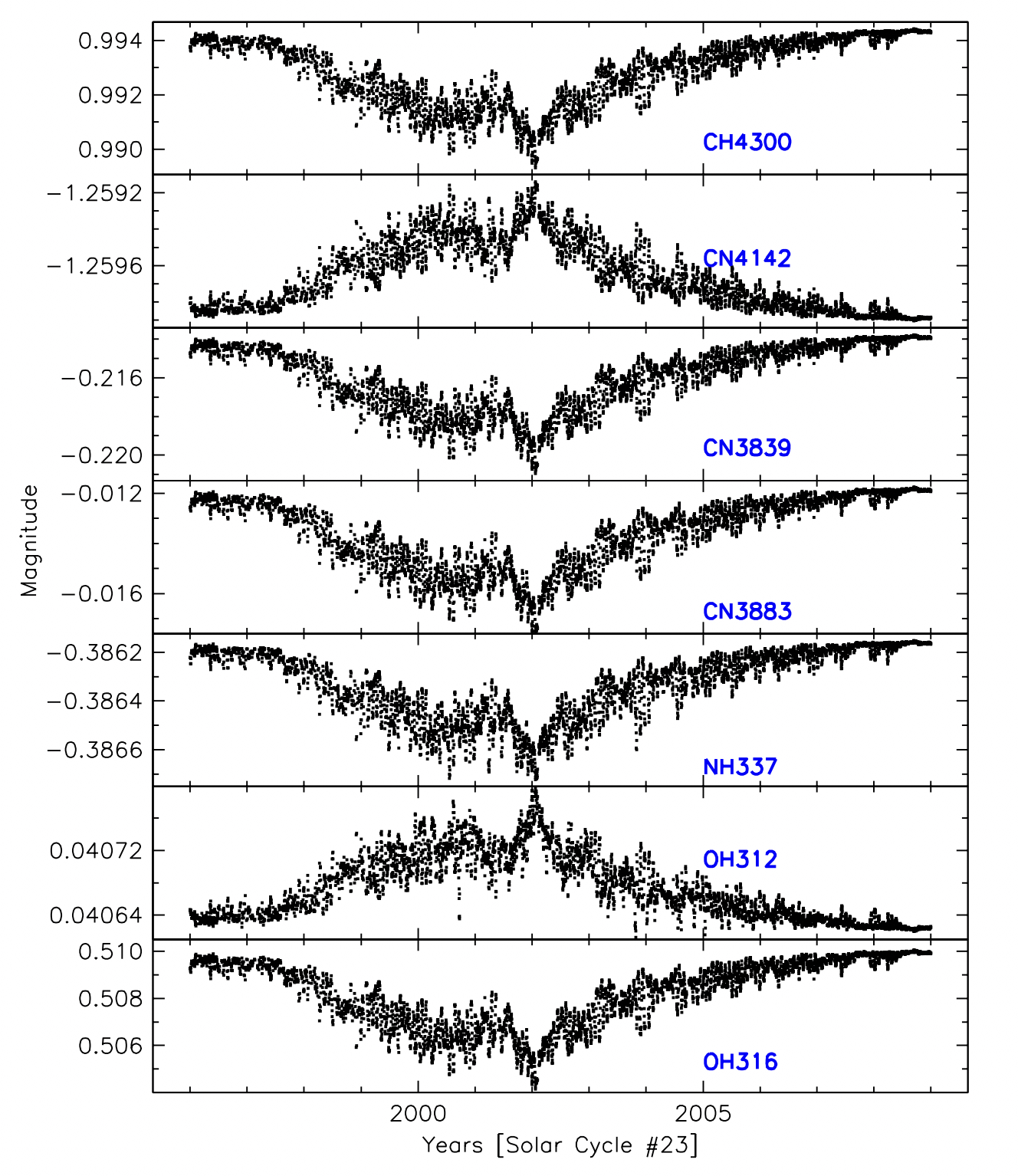} 
\caption{Illustration of the dependency of  seven   molecular
    indices on the Sun's $23^{\mathrm{rd}}$ magnetic cycle.
The $Y$-axis shows the strength of the inferred molecular indices,
expressed in magnitudes.
}
\label{f_7ind}
\end{figure}

Figure~\ref{f_7ind}  demonstrates  that  all seven  molecular  indices
exhibit   varying    degrees   of   sensitivity   to    the   opposing
${\sim}$$+$$1000$\,K                                               and
$-$$1000$\,K  temperature excursions  induced by  surface faculae  and
spots, respectively, and by magnetism more broadly.
Figure~\ref{f_7ind}  also reveals  important clues  about the
  molecular bands' physical nature and diagnostic limitations.
The most  striking  is the  apparent contradiction  between the
trends  traced  by  the  CN4142   and  CN3883  indices.  Notably,  the
observation  of such  divergent behaviours  within the  same star---the
Sun---enables us to  confidently rule out confounding  factors such as
interstellar reddening, chemical abundance variations, or instrumental
effects.
Rather, this  solar CN discrepancy  is likely rooted  in two
principal causes:

\begin{itemize}

\item Pseudo-continuum effects: The CN3883 region sits on the
  steep   bluewards   wing    of   the   broad   Ca\,\textsc{ii}\,H$+$K
  lines.  Intense line-blending  from  CH, NH,  and various  iron-peak
  elements creates  a `pseudo-continuum depression' that  can dilute
  the apparent strength  of the CN3883 index. In  contrast, the region
  around  $4142$\,\AA\  is spectrally  flatter,  allowing  for a  more
  robust and stable continuum definition.

\item  Formation   Physics:  These  indices   probe  distinct
  electronic transitions  and form  at different  depths in  the solar
  atmosphere. The  CN4142 band forms  deeper in the  photosphere where
  higher  gas  pressures  and densities  enhance  molecular  stability
  despite higher  temperatures.  Conversely,  CN3883 forms  in higher,
  more rarefied layers where the continuum is less well-defined.

\end{itemize}

A comparable discrepancy arises in  the OH312 and OH316 indices, where
subtle variations  in window definition and  continuum placement yield
markedly   different   cycle-modulated    responses.   As   with   the
CN4142--CN3883  pair,   this  reinforces  that   such  index-dependent
divergences stem  from local  atmospheric stratification  and bandpass
design rather than intrinsic abundance variations.

In conclusion, solar  molecular indices offer a  profound insight into
the interplay  between magnetic activity,  atmospheric stratification,
and  spectral   variability  \citep{shapiro2015}.  Figure~\ref{f_7ind}
demonstrates that while
  all solar  molecular indices  track the  magnetic field  cycle,
their  diagnostic consistency  is highly  dependent on  their physical
formation conditions.
Importantly, these  inconsistencies do  not reflect variations  in the
Sun's  elemental  abundances,  but  rather  arise  from  the  inherent
physical response  of the photosphere  to the magnetic cycle.
While  a full  extrapolation  of  these solar  trends  to the  diverse
temperatures  and metallicities  of  stars in  GCs  remains a  complex
challenge, the  solar evidence provides a  clear proof-of-concept that
the role of surface magnetism could be highly significant \citep[e.g.  
the NH response in Fig.~5 of][Fig.~12 of \citealt{milone2015}; and 
Fig.~1 of \citealt{lee2019}]{sbordone2011}.
Overall,  the  \textsc{MAGISTER  FAS}  framework  can,  in  principle,
account  for the  differential photometric  brightenings and  dimmings
observed in 2P and 1P populations as a natural observational signature
of surface magnetism.

\section{The  impact  of dynamical  encounters  in
  dense clusters  on  surface magnetism}
\label{s_dynamics}

To fully appreciate  the viability and appeal  of the \textsc{MAGISTER
  FAS} framework  from a dynamical standpoint,  one must contextualise
the extreme environmental conditions  characteristic of dense GCs.
Throughout this work, whenever a template was required to relate solar
magnetic properties to stars within a GC environment, we have referred
to  NGC104 and  the average  photometric and  spectroscopic
properties of  its members.
Interestingly,  from  a  dynamical   perspective,  NGC104  is  equally
noteworthy  and representative  of  clusters hosting  MPs.  It is  the
fourth-most-massive  of  the  ${\sim}150$ Galactic  globular  clusters
\citep{baumgardt2018} and possesses the  highest collision rate of any
GC in the Milky Way \citep{lanzoni2010}.
This extreme environment is observationally mirrored by its hosting of the 
second-largest population of millisecond pulsars and close-binary X-ray 
sources in the Galaxy \citep[e.g.][]{pooley2003, heinke2005}.
Consequently,  its  members  experience  a  remarkably  high  rate  of
dynamical encounters. Indeed,  as a stark illustration  of the extreme
conditions within NGC104, classic estimates suggest that up to tens of
per cent of the low-mass population in  a GC core may undergo a direct
physical collision \citep{hills1976}.

Nevertheless, a recent high-precision spectroscopic and photometric
analysis of NGC104 by \citet{mullerhorn2025} reveals an exceptionally
low total binary frequency of only $2.4$$\pm$$1.0$$\%$
\footnote{This NGC104-specific binary  frequency aligns with
      the  systematic  monitoring of  $968$  giant  stars across  $10$
      Galactic  GCs by \cite{lucatello2015},  who  inferred a  global
      binary                        frequency                       of
      $2.2$$\pm$$0.5$$\%$. Photometric data \citep{milone2012} further
      corroborate a  significantly lower total binary  fraction in GCs
      than in the Galactic field.}.
Even more striking is their estimate of the NGC104 binary
frequency among its Blue Straggler Stars (BSS)---the hallmarks
of stellar interaction.
While the formation of BSS is traditionally attributed to either
stellar mergers induced by direct collisions or mass-transfer
processes in binary systems \citep{sills2009, ferraro2009},
 \citet{mullerhorn2025} found the BSS binary fraction in NGC104 to be
only approximately three times higher than the cluster average.
This implies an NGC104 BSS-binary frequency
  ($\lesssim$$10$$\%$) that remains remarkably low compared to the Galactic
  field, where the binary fraction for solar-type dwarfs typically
  reaches ${\sim}50$$\%$ \citep{raghavan2010}.
This  GC  `binary  paradox'---the coexistence  of  extreme
density with low binarity---provides  two supporting arguments for the
\textsc{MAGISTER FAS} framework.

First,    this   suggests    that    the   vast    majority   of    GC
members---specifically  those  that   survived  without  merging---may
nonetheless  retain the  dynamical  imprints of  a  lifetime of  close
flybys and  grazing encounters, regardless of  their classification as
1P or 2P stars.

Second, while numerical and  empirical evidence suggest that dynamical
interactions---prevalent    in    high-density    environments    such as 
NGC104---can  effectively   trigger  enhanced  magnetic   activity  in
low-mass stars by facilitating the formation of rapidly rotating close
binaries \citep{pooley2003, heinke2010},  the resulting absolute field
strengths remain remarkably modest.
For example,  MS-MS collision models by  \citet{ryu2025} estimate that
the surface  magnetic fields  of thermally relaxed  collision products
increase,      but      only     to      values      between
$10$$\div$$10^4$\,Gauss.
Indeed, empirical confirmation of these estimates in the archetypal 
`icons' of stellar collisions---blue and yellow straggler stars---yields 
longitudinal magnetic fields on the order of a hundred to a few 
hundred Gauss \citep{hubrig2025}.
That  even   these  extreme   survivors---effectively  `post-merger'
remnants---exhibit such  relatively modest  field strengths  is highly
significant.
It  suggests a  fundamental upper  limit on  the global  mean
  field  strength  ($\langle$B$\rangle$)   even  when  generated  via
dynamical channels.
This  implies  that  the   surface  magnetic  intensity  envisaged  by
\textsc{MAGISTER  FAS} to  drive the  observed anomalies  need not  be
extreme.

Under this  framework, the MP  phenomenon would emerge as  the natural
manifestation of a cluster’s  specific dynamical history, manufactured
by  the  injection of  angular  momentum  during stochastic  dynamical
interactions.
Interestingly, evidence of this  increased angular momentum is readily
found   in  the   properties   of  a   significant  BY\,Dra   variable
population\footnote{BY\,Dra variables are typically K-type
  main-sequence   stars   \citep{chahal2022}   exhibiting   luminosity
  modulation  due  to  the   rotation  of  transiting  starspots.  The
  discovery   of   these   variables   in  NGC104   stems   from   the
  ground-breaking  HST  campaign  by \citet{gilliland2000}.  With  its
  $8.3$\,days   of  continuous   observation   and  high   photometric
  precision, this campaign remains the  closest GC equivalent to space
  missions     such     as    CoRoT     \citep{baglin2009},     Kepler
  \citep{borucki2010}, and TESS \citep{ricker2015}.} in NGC104.

Analysis  of  the   BY\,Dra  sample  in  NGC104   reveals  that  these
magnetically active variables exhibit colour and luminosity properties
virtually indistinguishable from the  presumably single (1P and/or 2P)
MS population \citep[cf. Fig.~20;][]{albrow2001}.
Out  of  the  $71$  variables  identified  by  \citet{albrow2001},  we
selected   a    sub-sample   of   $24$   stars    within   the   range
$18.3$$\le$m$_{\mathrm{F336W}}$$\le$$20.1$ to properly isolate true MS
dwarfs and exclude turn-off stars. This BY\,Dra subset exhibits a mean
rotation                           period                           of
${\sim}$$4.6$\,days,   a  value   highly  indicative   of  significant
rotational spin-up.
Indeed, for  a population  roughly twice  the solar  age, an  average rotation
period                                                              of
$\sim$$4.6$\,days—less  than  a  quarter  of  the  solar  equatorial  period
($P_{\rm
  rot,\odot}$$\simeq$$25.4$\,days)—stands  in  stark contradiction  to
the standard gyrochronological framework.
According  to the  classic  Skumanich relation  \citep{skumanich1972},
which predicts  that surface rotation  velocity decays as  the inverse
square  root  of  stellar  age,  these  BY\,Dra  variables  should  be
significantly  slower rotators  than  the Sun.   For  a population  of
$\sim$$10$--$12$\,Gyr,   one   would   expect  rotation   periods   in   the
$\sim$$35$--$40$\,day      range;      the     observed      average      of
$\sim$$4.6$\,days thus  represents a  profound discrepancy.
This discrepancy remains  striking even when compared to
recently     revised      weakened      magnetic     braking     models
\citep{vansaders2016}.
Such significantly shortened rotation  periods of the BY\,Dra
  variables  in NGC104  strongly  suggest a  heightened injection  of
angular momentum across the cluster population, providing the physical
engine for  the sustained  surface magnetic  activity proposed  by the
\textsc{MAGISTER FAS} framework.

This   cluster-specific,  environment-driven   injection  of   angular
momentum directly  echoes a fundamental  cosmic dichotomy, in  which a
heavily pronounced magnetic splitting  is potentially sustained within
massive stellar  cluster systems,  yet remains typically  absent among
field populations.

This architectural divide is once
  again deeply rooted  in the unique environmental  nature of globular
  clusters, which on average display an exceptionally low total binary
  fraction of only $\sim$$2.4$$\%$ as opposed to the
  $\sim$$50$$\%$ baseline typical of
  the Galactic field. 
  Indeed, as demonstrated by \citet{piotto2004}, the GCs' 
  `binary  paradox'   reflects  a  robust  and   highly  significant
  anti-correlation  between the  parent cluster's  total mass  and the
  relative  frequency  of  binary-mediated   products,  such  as  blue
  straggler stars.
  In  particular, the  cluster environment  of highly  massive systems
  such as NGC104  is fundamentally  hostile to soft  primordial binaries,
  systematically  disrupting them  or accelerating  their evolutionary
  depletion    compared   to    less    massive   globular    clusters
  \citep{piotto2004}, as  well as open  clusters or the  Galactic
  field \citep[cf.][Fig.~6.4]{momany2015}.
  Consequently,  the  observed  low  binary frequency  in  typical  GC
  systems  is not  an  indicator of  dynamical  passivity; rather,  it
  reflects  a continuous,  highly  aggressive  dynamical engine  where
  close grazing  encounters and tidal perturbations  constantly harass
  single stars and spin them up, ultimately driving heightened surface
  magnetic activity.

  Interestingly,  some empirical  validation for  the \textsc{MAGISTER
    FAS}  enhanced-rotation  hypothesis   has  recently  emerged  even
  outside the globular cluster domain.
  In a systematic  analysis of intermediate-age open clusters,
  \citet{pancino2018}  discovered   that  fast-rotating   dwarf  stars
  ($v\,\sin\,i$$\ge$$50$~km~s$^{-1}$) spontaneously  display remarkable
  bimodalities        in        $[\mathrm{Na}/\mathrm{Fe}]$        and
  $[\mathrm{O}/\mathrm{Fe}]$,      clear      $\mathrm{Na}$/$\mathrm{O}$
  anti-correlations,       and        extreme       depletions       in
  $[\mathrm{Mg}/\mathrm{Fe}]$  that  perfectly replicate  the  classic
  light-element  abundance patterns  observed in  complex GCs  such as
  NGC2808.
  Because  open  clusters  completely   lack  the  deep  gravitational
  potential  wells  or  massive  gas reservoirs  necessary  to  retain
  stellar  polluters or  sustain  successive star-formation  episodes,
  these chemical signatures cannot be  primordial or nuclear in origin.
  Instead, as argued by \citet{pancino2018},
  these anomalies emerge as an automated, superficial byproduct of the
  complex  non-nuclear  interplay  between rotational  mixing,  atomic
  diffusion, and  atmospheric transport mechanisms  operating directly
  on fast-rotating stellar surfaces.

We  note,  however, that  in  analysing  a relatively  larger
  sample from the Gaia-ESO  Survey, \citet{bragaglia2024} did not find
  clear  evidence of  these  $\mathrm{Na}$/$\mathrm{O}$ variations  in
  open clusters.
  Nonetheless, a  complementary perspective is provided  by wide-field
  Galactic    surveys     \citep[e.g.][]{horta2021,    fernandez2022,
    spite2022, leitinger2026}.
  For  instance,  in  their  analysis  of  the  APOGEE  DR17  dataset,
  \citet{fernandez2022}
  report the detection of $149$ N-rich field giants distributed
  across the Galactic bulge, metal-poor disc, and halo, all exhibiting
  a 2P-like nitrogen enrichment and carbon depletion
  anti-correlation.  These   stars  span   a  wide   metallicity  range
  ($-1.8$$<$$[\mathrm{Fe}/\mathrm{H}]$$<$$-0.7$) and show no strong
  signatures of binarity or pulsation.
  This field  population directly maps onto  the spatial distributions
  inferred from  APOGEE DR16 data by  \citet{horta2021}, who concluded
  that at  a distance of $1.5$\,kpc  from the Milky Way  centre, these
  2P-like  N-rich  stars  contribute  up  to
  $16.8^{+10.0}_{-7.0}\%$  of the  total  halo mass  budget, with  the
  fraction dropping  to $2.7^{+1.0}_{-0.8}\%$ out at  $10$\,kpc. 
  While  their low  intrinsic frequency  within the  field makes
  isolated detections inherently harder,  modern wide-area and all-sky
  spectroscopic surveys  are successfully  proving capable  of tracing
  statistically significant samples of  these 2P-like stars across the
  wider Galactic field.

Integrating these aforementioned lines of evidence regarding
  the sparse  but significant  detection of  2P-like stars  within the
  Galactic  field  and  open  clusters with  the  structural  boundary
  conditions    detailed   in    Appendix~\ref{s_young_clusters}—which
  demonstrate  that  the MP phenomenon  is  strictly
  governed by the intrinsic structural  or dynamical properties of the
  individual stars  themselves—offers key  insights into  the physical
  mechanisms  shaping the  field-versus-cluster  dichotomy across  all
  stellar regimes.
  In  isolated Galactic field settings  or  unperturbed  open clusters,  rapid
  stellar rotation is merely a brief, transient evolutionary phase; as
  stars  evolve  past  the  main-sequence turnoff  mass  threshold  of
  $\sim$$2$\,Gyr systems  and develop deep convective  envelopes, they
  rapidly shed  their angular momentum via  standard gyrochronological
  magnetic braking \citep{skumanich1972}.
  This smooth rotational 
  decay can be stochastically counteracted by the high primordial binarity 
  frequency  characteristic of  field  and  open cluster  environments
  ($\sim$$50$$\%$),
  which provides a distinct channel where tidal interactions can lock or 
  rejuvenate stellar rotation within close systems, thereby preserving the 
  localised surface activity envisaged by our framework.
  For the unperturbed 
  single-star population, however, these rotation-driven surface
  anomalies reported by
  \citet{pancino2018},  or the field  counterparts identified
  via   APOGEE  monitoring  \citep{fernandez2022}, are  quickly
  dissipated  and  become exceptionally  hard to  catch as  the 
  field population spins down and homogenises into a quiescent
  baseline.
  On  the other  hand, the  hyper-dense  environments of  massive
  systems such as NGC104 present a  relentless celestial traffic jam that
  subjects  convective members  to continuous  stochastic interactions
  over  a  Hubble time.
  This  persistent  environmental injection  of
  angular  momentum  effectively   halts  standard  magnetic  braking,
  permanently institutionalising a rapidly rotating, highly magnetised
  state  across a  substantial cohort  of low-mass  stars.

Thus, what
  manifests  as an  elusive, fleeting  superficial chemical  curiosity
  among   young  stars   in  open   clusters  and   unperturbed  field
  environments seems instead to be  dynamically locked in place within
  globular clusters  over cosmic  timescales, sustaining  a permanent,
  bimodal split into a pristine, quiescent baseline (1P) and a
  dynamically spun-up, magnetically active cohort (2P).

Remarkably,   the  Galactic   field  provides   a  compelling
  empirical  parallel to  our  expectation of  a fundamental  magnetic
  dichotomy among GC stars.
  A particularly  instructive baseline is provided  by the homogeneous
  analysis of over
  $\sim$$1600$ F/G/K main-sequence Galactic field stars extracted from
  the HARPS archive \citep{gomes2021}, which directly correlated their
  chromospheric
  $\mathrm{Ca}\,\textsc{ii}$\,$\mathrm{H}$$+$$\mathrm{K}$  emission (a
  direct proxy for surface  magnetic activity) with stellar parameters
  and evolutionary ages.
Their results robustly confirm long-standing configurations 
pioneered by \citet{henry1996}:
main-sequence field stars  can be broadly disentangled  into two major
sub-populations  consisting  of  magnetically   active   and  
  inactive  stars.
The emergence of  a natural magnetic dichotomy  among solar-like stars
is far  from trivial, especially  considering that the  Galactic field
dwarf sample  \citep{gomes2021} represents a highly  composite dataset
in both age and metallicity,  completely lacking the nearly monolithic
chemical and temporal distributions characteristic of coeval GC stars.
Yet, despite this underlying cosmic scatter, active and inactive field cohorts 
are cleanly distinguished by their corrected chromospheric emission ratios 
($\log R^{\prime}_{\rm HK}$), with active field dwarfs systematically exhibiting 
higher effective temperatures ($\Delta T_{\rm eff}$$\sim$$50\text{--}70$\,K) 
than their quiescent counterparts. This temperature offset seamlessly mimics 
the characteristic bluer, hotter trend that would define the active 2P sequence  
as envisaged by \textsc{MAGISTER FAS}.

\end{appendix} 

\end{document}